\documentclass[conference,comsoc,a4paper]{IEEEtran}
\IEEEoverridecommandlockouts
\usepackage{cite}
\usepackage{amsmath,amssymb,amsfonts}
\usepackage[nolist]{acronym}
\usepackage{algorithmic}
\usepackage{enumitem}
\usepackage{scalerel}
\usepackage{graphicx}
\usepackage{textcomp}
\usepackage{xcolor}
\usepackage{tikz}
\usepackage{bm}

\usetikzlibrary{calc}
\usetikzlibrary{positioning}
\usetikzlibrary{shapes.geometric}
\usetikzlibrary{arrows}
\usepackage{pgfplots}
\usepgfplotslibrary{groupplots}
\usepgfplotslibrary{colormaps}

\usepackage{float}
\pgfplotsset{compat=newest}

\pgfplotsset{
  colormap/twilight/.style={colormap={twilight}{[1pt]
  rgb(0pt)=(0.8857501584075443, 0.8500092494306783, 0.8879736506427196);
  rgb(25pt)=(0.38407269378943537, 0.46139018782416635, 0.7309466543290268);
  rgb(50pt)=(0.18488035509396164, 0.07942573027972388, 0.21307651648984993);
  rgb(75pt)=(0.6980608153581771, 0.3382897632604862, 0.3220747885521809);
  rgb(100pt)=(0.8857115512284565, 0.8500218611585632, 0.8857253899008712);
}}}

\makeatletter
\let\MYcaption\@makecaption
\makeatother

\usepackage[font=footnotesize]{subcaption}

\makeatletter
\let\@makecaption\MYcaption
\makeatother

\tikzset{>=latex}

\begin{acronym}
 \acro{R2M}{raw 2\textsuperscript{nd} moment}
 \acro{CSI}{channel state information}
 \acro{UE}{user equipment}
 \acro{EM}{electromagnetic}
 \acro{UL}{uplink}
 \acro{BS}{base station}
 \acro{TDD}{time division duplex}
 \acro{FDD}{frequency division duplex}
 \acro{ECC}{error-correcting code}
 \acro{MLD}{maximum likelihood decoding}
 \acro{HDD}{hard decision decoding}
 \acro{IF}{intermediate frequency}
 \acro{RF}{radio frequency}
 \acro{SDD}{soft decision decoding}
 \acro{NND}{neural network decoding}
 \acro{CNN}{convolutional neural network}
 \acro{ML}{maximum likelihood}
 \acro{GPU}{graphical processing unit}
 \acro{BP}{belief propagation}
 \acro{LTE}{Long Term Evolution}
 \acro{BER}{bit error rate}
 \acro{SNR}{signal-to-noise-ratio}
 \acro{ReLU}{rectified linear unit}
 \acro{BPSK}{binary phase shift keying}
 \acro{QPSK}{quadrature phase shift keying}
 \acro{AWGN}{additive white Gaussian noise}
 \acro{MSE}{mean squared error}
 \acro{LLR}{log-likelihood ratio}
 \acro{MAP}{maximum a posteriori}
 \acro{NVE}{normalized validation error}
 \acro{BCE}{binary cross-entropy}
 \acro{CE}{cross-entropy}
 \acro{BLER}{block error rate}
 \acro{SQR}{signal-to-quantisation-noise-ratio}
 \acro{MIMO}{multiple-input multiple-output}
 \acro{mMIMO}{massive multiple-input multiple-output}
 \acro{OFDM}{orthogonal frequency division multiplex}
 \acro{RF}{radio frequency}
 \acro{LOS}{line of sight}
 \acro{NLoS}{non-line of sight}
 \acro{NMSE}{normalized mean squared error}
 \acro{CFO}{carrier frequency offset}
 \acro{SFO}{sampling frequency offset}
 \acro{IPS}{indoor positioning system}
 \acro{TRIPS}{time-reversal IPS}
 \acro{RSSI}{received signal strength indicator}
 \acro{MIMO}{multiple-input multiple-output}
 \acro{ENoB}{effective number of bits}
 \acro{AGC}{automatic gain control}
 \acro{ADC}{analog to digital converter}
 \acro{ADCs}{analog to digital converters}
 \acro{FB}{front bandpass}
 \acro{FPGA}{field programmable gate array}
 \acro{JSDM}{Joint Spatial Division and Multiplexing}
 \acro{NN}{neural network}
 \acro{IF}{intermediate frequency}
 \acro{LoS}{line-of-sight}
 \acro{NLoS}{non-line-of-sight}
 \acro{DSP}{digital signal processing}
 \acro{AFE}{analog front end}
 \acro{SQNR}{signal-to-quantisation-noise-ratio}
 \acro{SINR}{signal-to-interference-noise-ratio}
 \acro{ENoB}{effective number of bits}
 \acro{PCB}{printed circuit board}
 \acro{EVM}{error vector mangnitude}
 \acro{CDF}{cumulative distribution function}
 \acro{MRC}{maximum ratio combining}
 \acro{MRP}{maximum ratio precoding}
 \acro{MRT}{maximum ratio transmission}
 \acro{DeepL}{deep-learning}
 \acro{DL}{downlink}
 \acro{SISO}{single-input single-output}
 \acro{SGD}{stochastic gradient descent}
 \acro{CP}{cyclic prefix}
 \acro{MISO}{Multiple Input Single Output}
 \acro{LMMSE}{linear minimum mean square error}
 \acro{ZF}{zero forcing}
 \acro{USRP}{universal software radio peripheral}
 \acro{RNN}{recurrent neural network}
 \acro{GRU}{gated recurrent unit}
 \acro{LSTM}{long short-term memory}
 \acro{NTM}{neural turing machine}
 \acro{DNC}{differentiable neural computer}
 \acro{TCN}{temporal convolutional network}
 \acro{FCL}{fully connected layer}
 \acro{MANN}{memory augmented neural network}
 \acro{RNN}{recurrent neural network}
 \acro{DNN}{deep neural network}
 \acro{FIR}{finite impulse response}
 \acro{BPTT}{back-propagation through time}
 \acro{GAN}{generative adversarial network}
 \acro{ELU}{exponential linear unit}
 \acro{tanh}{hyperbolic tangent}
 \acro{BICM}{bit-interleaved coded modulation}
 \acro{OTA}{over-the-air}
 \acro{IM}{intensity modulation}
 \acro{DD}{direct detection}
 \acro{RL}{reinforcement learning}
 \acro{SDR}{software-defined radio}
 \acro{WGAN}{Wasserstein GAN}
 \acro{BMD}{bit-metric decoding}
 \acro{BMI}{bit-wise mutual information}
 \acro{LDPC}{low-density parity-check}
 \acro{IDD}{iterative demapping and decoding}
 \acro{JSD}{Jensen-Shannon distance}
 \acro{MMSE}{minimum mean square error}
 \acro{FFT}{fast Fourier transform}
 \acro{DFT}{discrete Fourier transform}
 \acro{IFFT}{inverse fast Fourier transform}
 \acro{QAM}{quadrature amplitude modulation}
 \acro{EMD}{earth mover's distance}
 \acro{TDL}{tapped delay line}
 \acro{KL}{Kullback-Leibler}
 \acro{PRACH}{physical random access channel}
 \acro{URLLC}{ultra-reliable low-latency communication}
 \acro{ANOMA}{asynchronous non-orthogonal multiple access}
 \acro{FEC}{forward error correction}
 \acro{PAPR}{peak-to-average power ratio}
 \acro{APP}{a posteriori probability}
 \acro{COTS}{commercial off-the-shelf}
 \acro{PLL}{phase locked loop}
 \acro{STO}{sampling time offset}
 \acro{SFO}{sampling frequency offset}
 \acro{CFO}{carrier frequency offset}
 \acro{CPO}{carrier phase offset}
 \acro{CSI}{channel state information}
 \acro{GNSS}{global navigation satellite system}
 \acro{ELAA}{extremely large aperture array}
 \acro{UE}{user equipment}
 \acro{DICHASUS}{\underline{Di}stributed \underline{Cha}nnel \underline{S}ounder by \underline{U}niversity of \underline{S}tuttgart}
 \acro{JCaS}{Joint Communication and Sensing}
 \acro{CT}{continuity}
 \acro{TW}{trustworthiness}
 \acro{KS}{Kruskal's stress}
 \acro{PCA}{principal component analysis}
 \acro{AoA}{angle of arrival}
 \acro{RD}{Rajski's distance}
 \acro{ADP}{angle-delay profile}
 \acro{MPC}{multipath component}
 \acro{CIR}{channel impulse response}
 \acro{MAE}{mean absolute error}
 \acro{MDS}{multidimensional scaling}
 \acro{t-SNE}{t-distributed stochastic neighbor embedding}
 \acro{SM}{Sammon's mapping}
 \acro{CIRA}{channel impulse response amplitude}
 \acro{CS}{cosine similarity}
 \acro{FCF}{forward charting function}
 \acro{CMD}{correlation matrix distance}
 \acro{NR}{New Radio}
 \acro{3GPP}{3rd Generation Partnership Project}
 \acro{QuaDRiGa}{Quasi Deterministic Radio Channel Generator}
 \acro{TR}{technical report}
 \acro{VAE}{variational autoencoder}
 \acro{GSCM}{geometry-based stochastic channel model}
 \acro{GP}{gradient penalty}
 \acro{LTI}{linear time-invariant}
 \acro{RMS}{root mean square}
 \acro{UPA}{uniform planar array}
 \acro{MUSIC}{MUltiple SIgnal Classification}
 \acro{PDP}{power delay profile}
 \acro{PDF}{probability density function}
\end{acronym}
\definecolor{mittelblau}{RGB}{0, 126, 198}
\definecolor{violettblau}{cmyk}{0.9, 0.6, 0, 0}
\definecolor{rot}{RGB}{238, 28 35}
\definecolor{apfelgruen}{RGB}{140, 198, 62}
\definecolor{gelb}{RGB}{1, 221, 0}
\definecolor{orange}{RGB}{244, 111, 33}
\definecolor{pink}{RGB}{237, 0, 140}
\definecolor{lila}{RGB}{128, 10, 145}
\definecolor{hellgrau}{RGB}{224, 224, 224}
\definecolor{mittelgrau}{RGB}{128, 128, 128}
\definecolor{dunkelgrau}{RGB}{80,80,80}
\definecolor{anthrazit}{RGB}{19, 31, 31}

\DeclareMathOperator*{\argmin}{arg\,min}

\usetikzlibrary{svg.path}

\def\BibTeX{{\rm B\kern-.05em{\sc i\kern-.025em b}\kern-.08em
    T\kern-.1667em\lower.7ex\hbox{E}\kern-.125emX}}

\definecolor{orcidlogocol}{HTML}{A6CE39}
\tikzset{
  orcidlogo/.pic={
    \fill[orcidlogocol] svg{M256,128c0,70.7-57.3,128-128,128C57.3,256,0,198.7,0,128C0,57.3,57.3,0,128,0C198.7,0,256,57.3,256,128z};
    \fill[white] svg{M86.3,186.2H70.9V79.1h15.4v48.4V186.2z}
                 svg{M108.9,79.1h41.6c39.6,0,57,28.3,57,53.6c0,27.5-21.5,53.6-56.8,53.6h-41.8V79.1z M124.3,172.4h24.5c34.9,0,42.9-26.5,42.9-39.7c0-21.5-13.7-39.7-43.7-39.7h-23.7V172.4z}
                 svg{M88.7,56.8c0,5.5-4.5,10.1-10.1,10.1c-5.6,0-10.1-4.6-10.1-10.1c0-5.6,4.5-10.1,10.1-10.1C84.2,46.7,88.7,51.3,88.7,56.8z};
  }
}

\newcommand\orcidicon[1]{
}

\begin{document}

\title{GAN-based Massive MIMO Channel Model Trained on Measured Data
\thanks{This work is supported by the German Federal Ministry of Education and Research (BMBF) within the projects Open6GHub (grant no. 16KISK019) and KOMSENS-6G (grant no. 16KISK113).}}
\author{\IEEEauthorblockN{Florian Euchner\textsuperscript{\orcidicon{0000-0002-8090-1188}}, Janina Sanzi, Marcus Henninger\textsuperscript{\orcidicon{0000-0003-2520-8943}}, Stephan ten Brink\textsuperscript{\orcidicon{0000-0003-1502-2571}} \\}

\IEEEauthorblockA{
Institute of Telecommunications, Pfaffenwaldring 47, University of  Stuttgart, 70569 Stuttgart, Germany \\ \{euchner,henninger,tenbrink\}@inue.uni-stuttgart.de
}
}

\newpage

\maketitle

\begin{abstract}
    Wireless channel models are a commonly used tool for the development of wireless telecommunication systems and standards.
    The currently prevailing geometry-based stochastic channel models (GSCMs) were manually specified for certain environments in a manual process requiring extensive domain knowledge, on the basis of channel measurement campaigns.
    By taking into account the stochastic distribution of certain channel properties like Rician k-factor, path loss or delay spread, they model the distribution of channel realizations.
    Instead of this manual process, a generative machine learning model like a generative adversarial network (GAN) may be used to automatically learn the distribution of channel statistics.
    Subsequently, the GAN's generator may be viewed as a channel model that can replace conventional stochastic or raytracer-based models.
    We propose a GAN architecture for a massive MIMO channel model, and train it on measurement data produced by a distributed massive MIMO channel sounder.
\end{abstract}


\section{Introduction}
Channel modeling is essential in communications engineering.
Fundamental theoretic results are often derived starting from some of the simplest channel models (e.g., memoryless discrete-time and discrete-value channels or \ac{AWGN} channels) and proposals for practical waveforms and modulation formats are motivated using simple stochastic channel models (e.g., Rician Fading, or Clarke's model \cite{clarke1968statistical}).
More involved channel models have been developed for researching practical wireless communication systems, that must be able to withstand various degrees of frequency- and/or time-selective fading, hardware impairments and other effects.
One such channel model was proposed by the \ac{3GPP} for the frequency range $0.5-100\,\mathrm{GHz}$ in \ac{TR} 38.901~\cite{5gchannelmodel}, which is an example for a so-called \ac{GSCM}.
These channel models define a set of stochastic parameters, including information about the distribution of delay spreads or path loss exponents, and guarantee some level of spatial coherence, for example by ensuring the spatial distribution follows some three-dimensional, correlated noise process.
While stochastic channel models usually only define the model (e.g., procedures, distributions, parameters), channel simulators like \ac{QuaDRiGa} \cite{burkhardt2014quadriga} may be used to produce channel realizations, for example in the form of \acp{CIR}.
For even better spatial consistency, raytracing-based channel simulators may be used as channel models \cite{sionna-rt}.
As evidenced by lots of successful, recent research based on stochastic or raytracer-based channel models, these existing solutions work well for the intended purposes.
However, in both cases, there is a lot of manual work required to come up with a model:
To obtain a stochastic channel model that describes a certain environment for a set of system parameters (e.g., carrier frequency), one first has to obtain channel measurements for the particular environment, to then identify relevant stochastic parameters and to finally extract these parameters from the measurements.
The inherent danger in this process is that relevant parameters may be overlooked or that the modeled environment simply does not match the system the model is used for.
Aforementioned \ac{TR} 38.901 channel model is only valid for a limited set of manually selected scenarios of interest, which define environments labeled, for example, ``urban macrocell'' or ``indoor factory''.

Raytracing-based channel models offer more flexibility to the user, but require significantly more effort related to the definition of the 3D model and material properties.
What is more, the user of a channel model may not care about specifics of one particular environment, which a raytracing-based channel model offers, but may want to rather evaluate some system on a certain type of environment.
While raytracing may be used to model an environment in a faithful, consistent way, stochastic channel models provide more variety and randomness in their channel realizations.
Finding a trade-off between faithfulness and randomness is a rather fundamental issue in channel modeling, and the right balance has to be determined depending on the particular application.

\begin{figure}
    \centering
    \scalebox{0.9}{
        \begin{tikzpicture}
            \node (dataset) at (-2, 0.8) [cylinder, aspect = 0.8, rotate = 90, draw, very thick, minimum width = 2.2cm, minimum height = 1.2cm] {};
            \node (generator) at (-2, -0.8) [draw, fill = mittelblau!10!white, very thick, minimum width = 2.2cm, minimum height = 1.2cm] {};
            
            \node [align = center] at (dataset) {Channel\\Dataset $\mathcal S_\mathrm{train}$};
            \node at (generator) {Generator $G$};
            
            \node (noiseinput) [left = 0.4cm of generator, align = center, yshift = 0.45cm] {Noise\\[-0.1cm]Vector $\mathbf n$};
            \node (posinput) [left = 0.4cm of generator, align = center, yshift = -0.45cm] {Desired\\[-0.1cm]Position $\mathbf x$};
            \draw [thick, -latex] (noiseinput) -- ($(generator.west) + (0, 0.45)$);
            \draw [thick, -latex] (posinput) -- ($(generator.west) + (0, -0.45)$);
            
            \node (critic) [draw, very thick, minimum width = 1.7cm, minimum height = 1.2cm, fill = mittelblau!10!white] at (2.2, 0) {Critic $C$};
            \draw [very thick, -o] (dataset.south) -- +(1.4, 0) node[midway, above] {$\mathbf H^{(l)}, \mathbf x^{(l)}$} coordinate (datasetout);
            \draw [very thick, -o] (generator.east) -- +(1.4, 0) node[midway, above] {$\mathbf {\hat H}, \mathbf x$}  coordinate (generatorout);
            
            \draw [thick, -latex, shorten <= 0.1cm] ($(datasetout) + (-0.125, 0)$) -- (critic.west);
            \draw [thick, dashed, -latex, shorten <= 0.1cm] ($(generatorout) + (-0.125, 0)$) -- (critic.west);
            
            \node (score) [right = 0.4cm of critic] {Score};
            \draw [thick, -latex] (critic) -- (score);
            
            \draw[latex-latex, dashed, thick, red!50!black] (critic.west)+(215:0.8cm) arc[start angle=215, end angle=145, radius=0.8cm];
    \end{tikzpicture}
    }
    \caption{Structural overview of the Wasserstein \ac{GAN} for channel modeling. The generator $G$ learns the \ac{CSI} distribution of the channel.}
    \label{fig:ganpicture}
\end{figure}
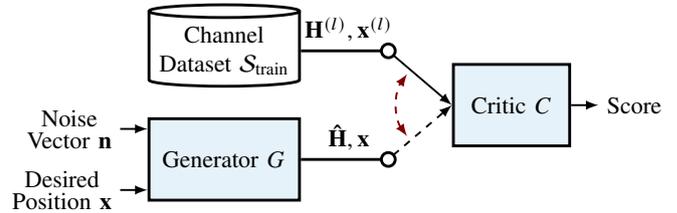

This work\footnote{\raggedright Partial source code is publicly available at: \texttt{https://github.com/Jeija/GAN-Wireless-Channel-Model}} analyzes a generative machine learning-based approach, which can automate the procedure of parametrizing a channel model on measurement data.
Instead of specifying relevant stochastic parameters manually, the generative channel model should automatically learn the distribution of the measured \ac{CSI}, based on samples from one or multiple datasets.
In particular, we explore the use of a \ac{GAN} \cite{goodfellow2014generative} (a \ac{WGAN} \cite{arjovsky2017wasserstein} with \ac{GP} \cite{gulrajani2017improved}, to be precise) for automated learning of stochastic channel models.
The basic structure of such a \ac{GAN}-based channel model is illustrated in Fig.~\ref{fig:ganpicture}.

\subsection{Related Work and Contributions}

\begin{figure}
    \centering
    \begin{subfigure}{0.9\columnwidth}
        \centering
        \scalebox{0.9}{
            \begin{tikzpicture}
                \node (channel) [draw, align = center, thick, minimum width = 1cm, minimum height = 1cm] at (0, 0) {Linear\\Channel};
                \node (input) [left = 2cm of channel, align = center, anchor  = center] {Channel Input\\$\mathbf s$};
                \node (output) [right = 2cm of channel, align = center, anchor  = center] {Channel Output\\$\mathbf y$};
                \node (gan) [draw, fill = mittelblau!10!white, thick, minimum width = 1cm, minimum height = 1cm, below = 1.2cm of channel] at (0, 0) {Generator $G$};
                
                \draw [-latex] (input)  -- (channel);
                \draw [-latex] (channel)  -- (output);
                \draw [-latex] (gan)  -- (channel) node[midway, right] {CSI $\mathbf H$};
            \end{tikzpicture}
        }
        \caption{Linear type: Generator produces instantaneous CSI}
    \label{fig:channelmodel-types-linear}
    \end{subfigure}
    
    \vspace{0.5cm}
    
    \begin{subfigure}{0.9\columnwidth}
        \centering
        \scalebox{0.9}{
            \begin{tikzpicture}
                \node (gan) [draw, fill = mittelblau!10!white, thick, minimum width = 1cm, minimum height = 1cm] at (0, 0) {Generator $G$};
                \node [left = 1.8cm of gan, align = center, anchor  = center] {Channel Input\\$\mathbf s$};
                \node [right = 1.8cm of gan, align = center, anchor  = center] {Channel Output\\$\mathbf y$};
                \draw [-latex] (input)  -- (gan);
                \draw [-latex] (gan)  -- (output);
            \end{tikzpicture}
        }
        \caption{Nonlinear type: Generator conditioned on channel input}
        \label{fig:channelmodel-types-nonlinear}
    \end{subfigure}
    \caption{Different types of GAN-based channel models.}
    \label{fig:channelmodel-types}
\end{figure}
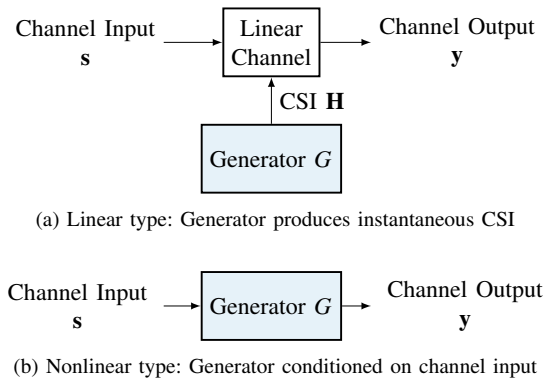

\begin{figure*}
    \centering
    \begin{subfigure}[b]{0.29\textwidth}
        \centering
        \includegraphics[width=0.9\textwidth, trim = 30 200 30 0, clip]{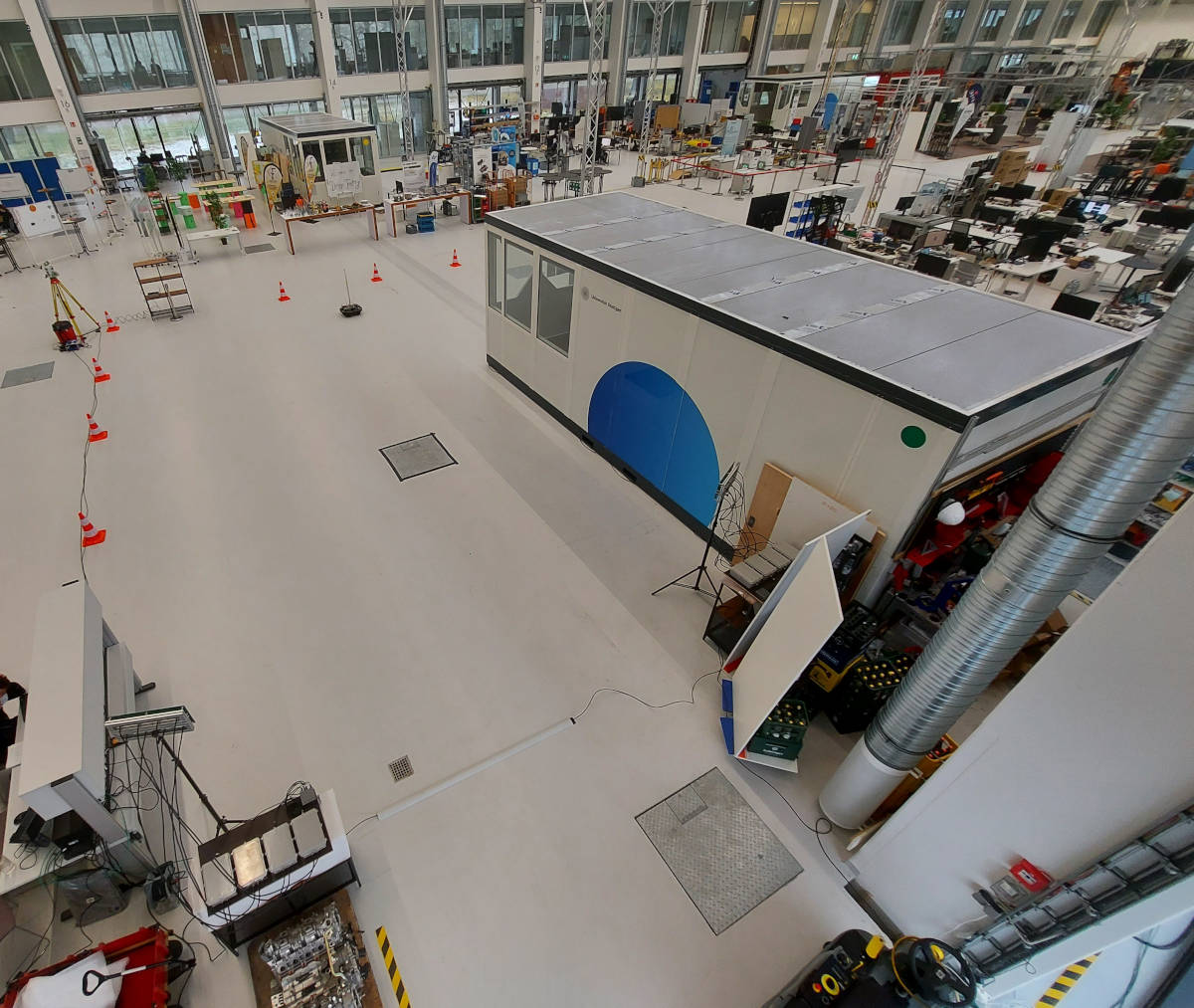}
        \vspace{0.8cm}
        \caption{}
    \end{subfigure}
    \begin{subfigure}[b]{0.3\textwidth}
        \centering
        \begin{tikzpicture}
            \begin{axis}[
                width=0.729\columnwidth,
                height=0.6\columnwidth,
                scale only axis,
                xmin=-15.5,
                xmax=6.1,
                ymin=-18.06,
                ymax=-0.5,
                xlabel = {Coordinate $\mathbf x_1 ~ [\mathrm{m}]$},
                ylabel = {Coordinate $\mathbf x_2 ~ [\mathrm{m}]$},
                ylabel shift = -8 pt,
                xlabel shift = -4 pt,
                xtick={-10, -6, -2, 2}
            ]
                \addplot[thick,blue] graphics[xmin=-14.5,ymin=-17.06,xmax=4.1,ymax=-0.5] {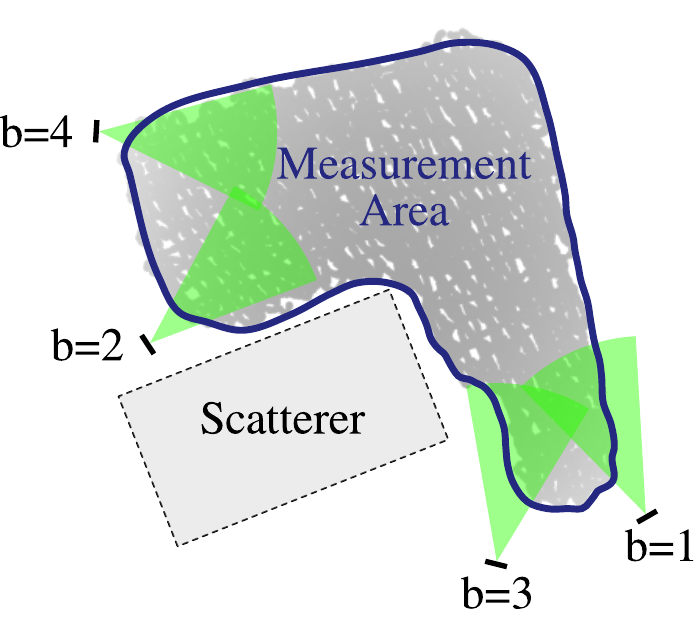};
            \end{axis}
        \end{tikzpicture}
        \vspace{-0.1cm}
        \caption{}
        \label{fig:labelled-area}
    \end{subfigure}
    \begin{subfigure}[b]{0.26\textwidth}
        \centering
        \begin{tikzpicture}
            \begin{axis}[
                width=0.7\columnwidth,
                height=0.7\columnwidth,
                scale only axis,
                xmin=-12.5,
                xmax=2.5,
                ymin=-15.5,
                ymax=-0.5,
                xlabel = {Coordinate $\mathbf x_1 ~ [\mathrm{m}]$},
                ylabel = {Coordinate $\mathbf x_2 ~ [\mathrm{m}]$},
                ylabel shift = -8 pt,
                xlabel shift = -4 pt,
                xtick={-10, -6, -2, 2}
            ]
                \addplot[thick,blue] graphics[xmin=-12.5,ymin=-15.5,xmax=2.5,ymax=-0.5] {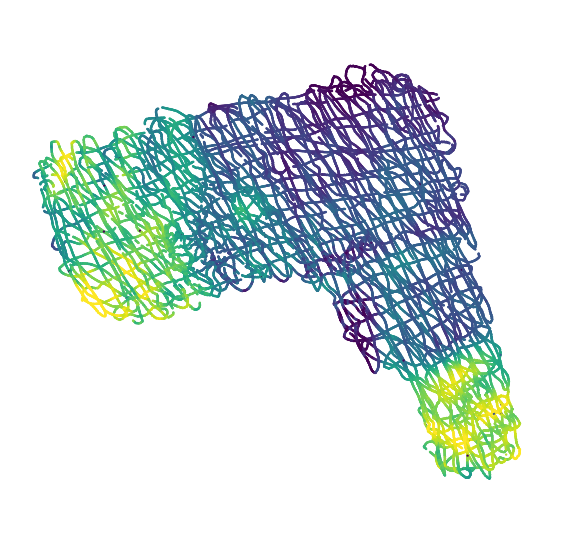};
                \node (tscirc) at (axis cs:-2,-6) [circle, draw = red, fill = white, opacity = 1, fill opacity = 0.5, very thick, inner sep=8pt, outer sep=0pt] {};
                \node [below = 0.05cm of tscirc, align = center, red, fill = white, opacity = 1.0, fill opacity = 0.5, text opacity = 1.0] {\footnotesize Training set\\[-0.1cm]\footnotesize cutout};
            \end{axis}
        \end{tikzpicture}
        \vspace{-0.1cm}
        \caption{}
        \label{fig:groundtruth-map}
    \end{subfigure}
    \begin{subfigure}[b]{0.09\textwidth}
        \begin{tikzpicture}
            \begin{axis}[
                hide axis,
                scale only axis,
                colormap/viridis,
                colorbar,
                point meta min=-10,
                point meta max=0,
                colorbar style={
                    height=4cm,
                    width=0.3cm,
                    ytick={-10,-5,0},
                    ylabel={Total RX Power $\lVert \mathbf H^{(l)} \rVert_\mathrm{F}^2$ in dB}
                }]
                \addplot [draw=none] coordinates {(0,0)};
            \end{axis}
        \end{tikzpicture}
        \vspace{0.3cm}
    \end{subfigure}
    \vspace{-0.3cm}
    \caption{Information about the environment the dataset was measured in: The figure shows (a) a photograph of the environment, (b) a top view map and (c) a scatter plot with datapoint positions $\mathbf x^{(l)}$ in $\mathcal S_\mathrm{meas}$, colorized with the total received power $\lVert \mathbf H^{(l)} \rVert_\mathrm{F}^2$. The antenna arrays in the map are drawn to scale as black rectangles and their directivity is indicated by the green sectors. The received power is normalized such that the maximum received power is $0\,\mathrm{dB}$.}
    \label{fig:industrial_environment}
    \vspace{-0.3cm}
\end{figure*}
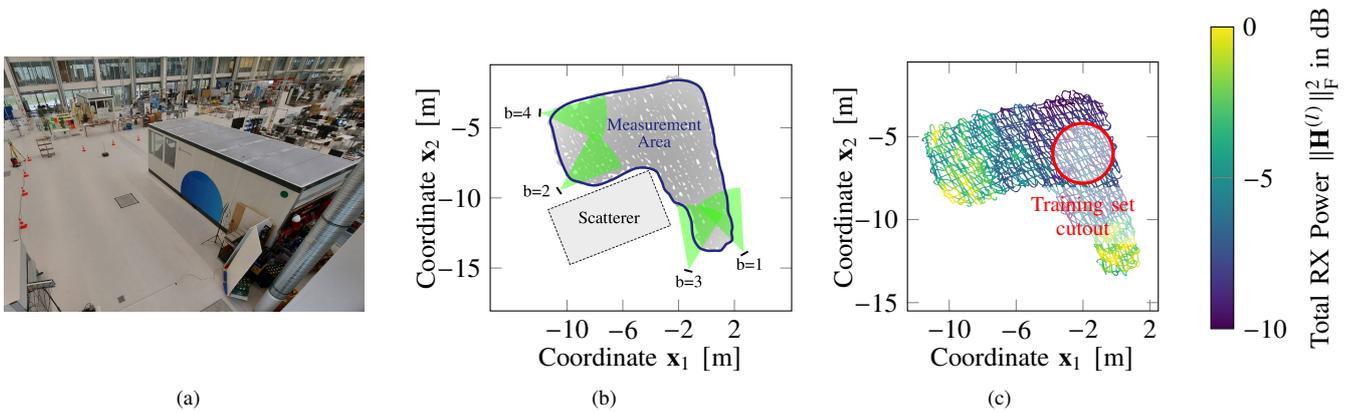

The idea of a \ac{GAN}-based channel model that learns the \ac{CSI} distribution, first proposed in \cite{yang2019generative}, has gained some attention in the community \cite{xiao2022channelgan, orekondy2022mimo, xie2023real, sengupta2023generative, juhava2023wireless}.
In \cite{yang2019generative}, the concept of \ac{GAN}-based channel modeling is introduced on a high level, motivating the idea by showing that a \ac{GAN} can learn the noise distribution of an \ac{AWGN} channel.
In \cite{xiao2022channelgan, orekondy2022mimo, xie2023real}, a \ac{GAN} is trained on channel realizations (impulse responses).
As an alternative to \acp{GAN}, \cite{sengupta2023generative} investigates the use of diffusion models for channel modeling and \cite{juhava2023wireless} compares a \ac{GAN}-based channel model to a \ac{VAE}-based model and a diffusion model.
In all previous research that the authors are familiar with, the generative model is trained on data that was itself produced by another channel model, usually related to aforementioned 3GPP \ac{TR} 38.901 \ac{GSCM}.

In a separate line of work, initiated by \cite{o2019approximating}, various authors have investigated using a conditional \ac{GAN} to learn the input-output relationship of the channel itself, which is a different type of \ac{GAN}-based channel model.
The distinction between \cite{yang2019generative} and \cite{o2019approximating}, which is also illustrated in Fig.~\ref{fig:channelmodel-types}, is the following: In \cite{yang2019generative}, the \ac{GAN}'s generator produces channel realizations described by \ac{CSI} in some representation (e.g., time-domain or frequency-domain), as in Fig.~\ref{fig:channelmodel-types-linear}.
On the other hand, in \cite{o2019approximating}, the generator \emph{becomes} the channel itself:
The generator is conditioned on the transmitted signal, and it must attempt to produce a realistic received signal, as in Fig.~\ref{fig:channelmodel-types-nonlinear}, which can be useful e.g. for end-to-end training \cite{dorner2020wgan}.
We restrict ourselves to the first, linear type of \ac{GAN}-based channel model as described in \cite{yang2019generative}.
As already outlined in \cite{orekondy2022mimo}, the advantages of generating \ac{CSI} as an intermediary product are the improved interpretability of the \ac{GAN}'s output and the better generalization to arbitrary input signals.
Furthermore, it ensures that the learned channel is indeed a \ac{LTI} system, whereas the approach in \cite{o2019approximating} can learn nonlinear channel models, which is undesirable for modeling physical propagation effects.

The architecture of our neural network draws inspiration from \cite{xiao2022channelgan}, being based on a \ac{WGAN} with \ac{GP}.
In contrast to all this previous work, however, we train our \ac{GAN}-based channel model on real-world data measured with a massive \ac{MIMO} channel sounder, as opposed to the previous efforts which were trained on another channel model.
In addition, we investigate the use of a conditional \acp{GAN} to generate channel realizations at defined locations, while ensuring some level of spatial consistency.

\subsection{Purpose of a GAN-based channel model}
The objective of a channel model depends on the application: A \ac{GSCM} may generate \ac{CSI} that follows a certain distribution, a raytracer may be used to predict realistic \ac{CSI}.
The purpose of \ac{GAN}-based channel models has not been well-defined in existing literature.
Therefore, we consider two different applications:
Using \ac{GAN}-based channel models as an alternative to \acp{GSCM} as a kind of \emph{random generator} for \ac{CSI} samples, or using \ac{GAN}-based channel models as a \emph{spatial inter- and extrapolator}, that should ideally predict measured \ac{CSI} at new locations.
While the use of a \ac{GAN} as an inter-/extrapolator has been suggested to us many times in conversations, we would like to remark that a different neural network structure, like an autoencoder, may be better-suited in that case.
In the following sections, we will analyze both types of applications and conclude that, in their current form, \ac{GAN}-based channel models of linear type are not particularly well-suited for either.

\section{System Model and Measurement Dataset}
Our objective is to model the distribution of a channel measurement dataset containing $L$ entries, that we refer to as \emph{datapoints} $l = 1, \ldots, L$.
We model the propagation channel between a distributed massive \ac{MIMO} antenna system, consisting of $B$ \acp{UPA} with half-wavelength spacing, $M_r$ rows and $M_c$ columns of antennas each, and a single-antenna \ac{UE}.
The channel is described by its complex-valued, time-domain \ac{CIR}.
The time-domain \ac{CSI} tensors can thus be written as $\mathbf H^{(l)} \in \mathbb C^{B \times M_r \times M_c \times N_\mathrm{tap}}$, where $N_\mathrm{tap}$ is the number of \ac{TDL} taps.
Furthermore, we assume that a \ac{UE} position $\mathbf x^{(l)} \in \mathbb R^2$ is known for every \ac{CSI} measurement.
Hence, one datapoint is a tuple constisting of \ac{CSI} tensor and \ac{UE} position, and the whole dataset may be written as the set

\[
    \mathcal S_\mathrm{meas} = \left\{ \left(\mathbf H^{(l)}, \mathbf x^{(l)} \right) \right\}_{l = 1, \ldots, L}.
\]

For this work, the dataset $\mathcal S_\mathrm{meas}$ was measured by \emph{\ac{DICHASUS}}, our distributed \ac{mMIMO} channel sounder, whose architecture is thoroughly described in \cite{dichasus2021}.
In brief, \ac{DICHASUS} is a massive \ac{MIMO}-\ac{OFDM} channel sounder.
It achieves long-term phase-coherence, even if antennas are distributed.
\ac{DICHASUS} provides large datasets containing \ac{OFDM} channel coefficients for all antennas and all subcarriers, along with side information like timestamps and accurate information about the positions of all antennas at all times.
While measurements are performed in frequency domain, it is only a matter of computing the Fourier transform to obtain the time-domain representation $\mathbf H^{(l)}$.

The considered dataset $\mathcal S_\mathrm{meas}$, containing $L = 83893$ datapoints, is a subset of \emph{dichasus-cf0x} \cite{dataset-dichasus-cf0x}, and was captured in an industrial environment with $B = 4$ separate antenna arrays made up of $M = 2 \times 4 = 8$ antennas each.
A total of $N_\mathrm{sub} = 1024$ \ac{OFDM} channel coefficients were measured at a carrier frequency of $1.272\,\mathrm{GHz}$ and with a channel bandwidth of $50\,\mathrm{MHz}$, though we restrict ourselves to $N_\mathrm{tap} = 48$ time-domain taps for all subsequent analyses.
The single dipole transmit antenna is mounted on top of a robot, which travels along a set of trajectories inside a defined, L-shaped area, with an overall size of approximately $14\,\mathrm{m} \times 14\,\mathrm{m}$.
A photo and a top view map of the environment are shown in Fig.~\ref{fig:industrial_environment}.
A large metal container is located at the inner corner of the L-shape, blocking the \ac{LoS}.
The datapoint positions $\mathbf x^{(l)}$ are shown in Fig.~\ref{fig:industrial_environment}.

We define training set $\mathcal S_\mathrm{train} \subset \mathcal S_\mathrm{meas}$ and test set $\mathcal S_\mathrm{test}  \subset \mathcal S_\mathrm{meas}$ as two disjoint subsets of the complete measurement dataset $\mathcal S_\mathrm{meas}$.
The training set contains $|\mathcal S_\mathrm{train}| = 17857$ datapoints, the test set contains $|\mathcal S_\mathrm{test}| = 20973$ datapoints.
The training set is used to train the \ac{GAN} or to parametrize the baselines.
The \ac{GAN} (or a baseline) then generates \ac{CSI} at the test set locations, and the generated \ac{CSI} is compared to the true, measured \ac{CSI}.
The test set is obtained from $\mathcal S_\mathrm{meas}$ by choosing every fourth datapoint.
We do the same for the training set, except with an offset, and then additionally cut a "hole" with a diameter of $4\,\mathrm{m}$ into the training set (as visible in Fig.~\ref{fig:groundtruth-map} and Fig.~\ref{fig:ds-trainingset}).
This hole is meant to test the ability of the \ac{GAN} and baselines to interpolate \ac{CSI} at previously unseen locations.

\section{GAN-Based Channel Model}
Fig.~\ref{fig:ganpicture} shows the well-known conditional \ac{WGAN} training setup:
For the desired position $\mathbf x$ (the condition) and as a function of some noise vector $\mathbf n = (n_1, \ldots, n_K)$, $n_k \sim \mathcal N(0, 1)$, $K = 128$, the generator $G(\mathbf x, \mathbf n)$ produces a \emph{generated} channel realization $\mathbf { \hat H } \in \mathbb C^{B \times M_r \times M_c \times N_\mathrm{tap}}$.
As its input, the critic $C$ either receives a \emph{real} (i.e., measured) channel $\mathbf H^{(l)}$ from $\mathcal S_\mathrm{train}$, or a generated channel $\mathbf {\hat H}$, always in conjunction with the condition.
The critic's objective is to detect whether the input it receives is real or fake, by assigning a score reflective of the \emph{realness} of the input sample.
We train the generator and critic to operate on the \emph{complete} \ac{CSI} tensors $\mathbf H^{(l)}$, which contain time-domain tap coefficients for all $B \times M_r \times M_c$ anntennas and all taps.

\begin{figure}
    \centering
    \begin{subfigure}{0.9\columnwidth}
        \centering
        \scalebox{0.82}{
            \begin{tikzpicture}
                \node (layer0) [rectangle, draw, minimum width = 5cm, minimum height = 0.5cm] at (0, 0) {Dense, 512 Neurons, ReLU};
                \node (layer1) [below = 0.3cm of layer0, rectangle, draw, minimum width = 5cm, minimum height = 0.5cm] {Dense, 512 Neurons, ReLU};
                \node (layer2) [below = 0.3cm of layer1, rectangle, draw, minimum width = 5cm, minimum height = 0.5cm] {Dense, 1024 Neurons, ReLU};
                \node (layer3) [below = 0.3cm of layer2, rectangle, draw, minimum width = 5cm, minimum height = 0.5cm] {Dense, 2048 Neurons, ReLU};
                \node (layer4) [below = 0.3cm of layer3, rectangle, draw, minimum width = 5cm, minimum height = 0.5cm] {Dense, 3072 Neurons, linear};
    
                \draw [-latex] (layer0) -- (layer1);
                \draw [-latex] (layer1) -- (layer2);
                \draw [-latex] (layer2) -- (layer3);
                \draw [-latex] (layer3) -- (layer4);
                
                \node (noise) [align = center] at ($(layer0.north) + (-1.2, 0.5)$) {$\mathbf n \in \mathbb R^K$};
                \node (pos) [align = center] at ($(layer0.north) + (1.2, 0.5)$) {Position $\mathbf x \in \mathbb R^2$};
                
                \draw [-latex] (noise) -- (noise |- layer0.north);
                \draw [-latex] (pos) -- (pos |- layer0.north);
                
                \node (output) [align = center] at ($(layer4.south) + (0, -0.7)$) {CSI $\mathbf {\hat H} \in \mathbb C^{B \times M_r \times M_c \times N_\mathrm{tap}}$};
                \draw [-latex] (layer4) -- (output);
            \end{tikzpicture}
        }
        \caption{Structure of generator neural network $G$, with random noise input $\mathbf n$ and condition $\mathbf x$ (position)}
    \label{fig:layers-generator}
    \end{subfigure}
    
    \vspace{0.5cm}
    
    \begin{subfigure}{0.9\columnwidth}
        \centering
        \scalebox{0.82}{
            \begin{tikzpicture}
                \node (layer0) [rectangle, draw, minimum width = 5cm, minimum height = 0.5cm] at (0, 0) {Dense, 160 Neurons, ReLU};
                \node (layer1) [below = 0.3cm of layer0, rectangle, draw, minimum width = 5cm, minimum height = 0.5cm] {Dense, 100 Neurons, ReLU};
                \node (layer2) [below = 0.3cm of layer1, rectangle, draw, minimum width = 5cm, minimum height = 0.5cm] {Dense, 50 Neurons, ReLU};
                \node (layer3) [below = 0.3cm of layer2, xshift = -2cm, rectangle, draw, minimum width = 9cm, minimum height = 0.5cm] {Dense, 20 Neurons, ReLU};
                \node (layer4) [below = 0.3cm of layer3, rectangle, draw, minimum width = 9cm, minimum height = 0.5cm] {Dense, 10 Neurons, ReLU};
                \node (layer5) [below = 0.3cm of layer4, rectangle, draw, minimum width = 9cm, minimum height = 0.5cm] {Dense, 1 Neuron, linear};
    
                \draw [-latex] (layer0) -- (layer1);
                \draw [-latex] (layer1) -- (layer2);
                \draw [-latex] (layer2) -- (layer2 |- layer3.north);
                \draw [-latex] (layer3) -- (layer4);
                \draw [-latex] (layer4) -- (layer5);
                
                \node (csi) [align = center] at ($(layer0.north) + (0.5, 0.5)$) {CSI $\mathbf {\hat H} \in \mathbb C^{B \times M_r \times M_c \times N_\mathrm{tap}}$};
                
                \node (ds) [align = center] at ($(layer0.north) + (-3, 0.5)$) {$\hat { \mathbf{DS} }_\mathrm{rms} \in \mathbb C^{B \times M_r \times M_c}$};
                
                \node (pos) [align = center] at ($(layer0.north) + (-5.75, 0.5)$) {Position $\mathbf x \in \mathbb R^2$};

                \draw [-latex] (csi) -- (csi |- layer0.north);
                \draw [-latex] (ds.south) -- (ds.south |- layer3.north);
                \draw [-latex] (pos.south) -- (pos.south |- layer3.north);
                
                \node (output) [align = center] at ($(layer5.south) + (0, -0.7)$) {Critic score $C(\mathbf { \hat H }) \in \mathbb R$};
                \draw [-latex] (layer5) -- (output);
            \end{tikzpicture}
        }
        \caption{Structure of critic neural network $C$, where input $\hat { \mathbf{DS} } \in \mathbb C^{B \times M_r \times M_c}$ denotes the normalized delay spread computed from $\hat {\mathbf H}$}
        \label{fig:layers-critic}
    \end{subfigure}
    \caption{Structure of neural networks. Reshaping layers and the conversion from complex representation to separate real / imaginary parts are omitted.}
    \label{fig:layers}
    \vspace{-0.4cm}
\end{figure}
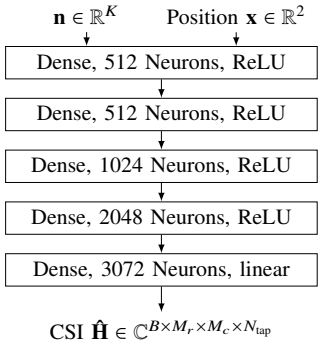
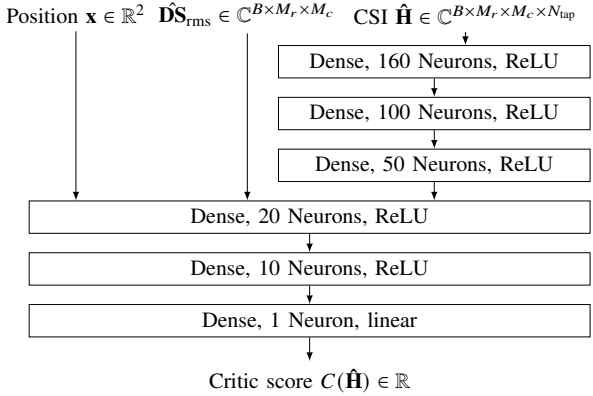

Generator and critic are dense neural networks, structured as shown in Fig.~\ref{fig:layers-generator} and Fig.~\ref{fig:layers-critic}.
In the neural network, the complex-valued \ac{CSI} is processed in separate real / imaginary parts, and conversion and reshaping layers are omitted from Fig.~\ref{fig:layers-generator} and Fig.~\ref{fig:layers-critic} for the sake of brevity.
In addition to the raw generated \ac{CSI} $\hat { \mathbf H }$, the critic is also provided antenna-specific \ac{RMS} delay spread values derived from $\hat { \mathbf H }$, the computation of which is explained in Sec. \ref{sec:delayspread}.
The behind providing \ac{RMS} delay spread values is that they may aid the critic in scoring the realness of a generated sample.
The \ac{WGAN} is trained with \ac{GP} \cite{gulrajani2017improved} with \ac{GP} coefficient $\lambda = 10$.
Delay spread values and the conditional positions $\mathbf x$ are normalized such that all numbers are in range $[-1, 1]$.
The \ac{CSI}, except for being split into real and imaginary parts, does not undergo any additional treatment like normalization.

\section{Linear Interpolation Baseline}
To judge whether a \ac{GAN}-based channel model is useful as an \emph{interpolator}, we compare it to a simple classical interpolation baseline.
While more advanced interpolation methods like, for example, Kriging are well-known, we employ a simpler linear interpolation baseline:
We construct the interpolant $F_\mathrm{interp}: \mathbb R^2 \to \mathbb C^{B\times M_r \times M_c \times N_\mathrm{tap}}$ by first computing the Delaunay triangulation of the two-dimensional training set positions, and then performing linear barycentric \ac{CSI} interpolation on each triangle as follows:
For a single triangle, let $\mathbf s \in \mathbb R^3$ denote the barycentric coordinate vector such that $\sum_i \mathbf s_i = 1$ and let $\mathbf H^{(\Delta 1)}$, $\mathbf H^{(\Delta 2)}$, $\mathbf H^{(\Delta 3)}$ denote the \ac{CSI} tensors of the corresponding vertices of the triangle.
To account for unknown global starting phases $\bm \varphi \in \mathbb R^3$ in the \ac{CSI} tensors, we find the interpolated \ac{CSI} tensor $\hat {\mathbf H}$ as the solution to the optimization problem
\[
    (\hat {\mathbf H}, \bm \varphi) = \argmin_{(\mathbf H, \bm \varphi)} \sum_{i=1}^3 s_i \left\lVert \mathbf H^{(\Delta i)} - \mathrm e^{\mathrm j \bm \varphi_i} \mathbf H \right\rVert_\mathrm{2}^2,
\]
where $\lVert \cdot \rVert_2$ is the straightforward generalization of the Euclidean norm to tensors and subtraction is applied elementwise.
The optimization problem is solved by separate differentiation with respect to $\bm \varphi$ and $\mathbf H$, which yields a simple coordinate descent interpolation algorithm.

\section{Evaluation}
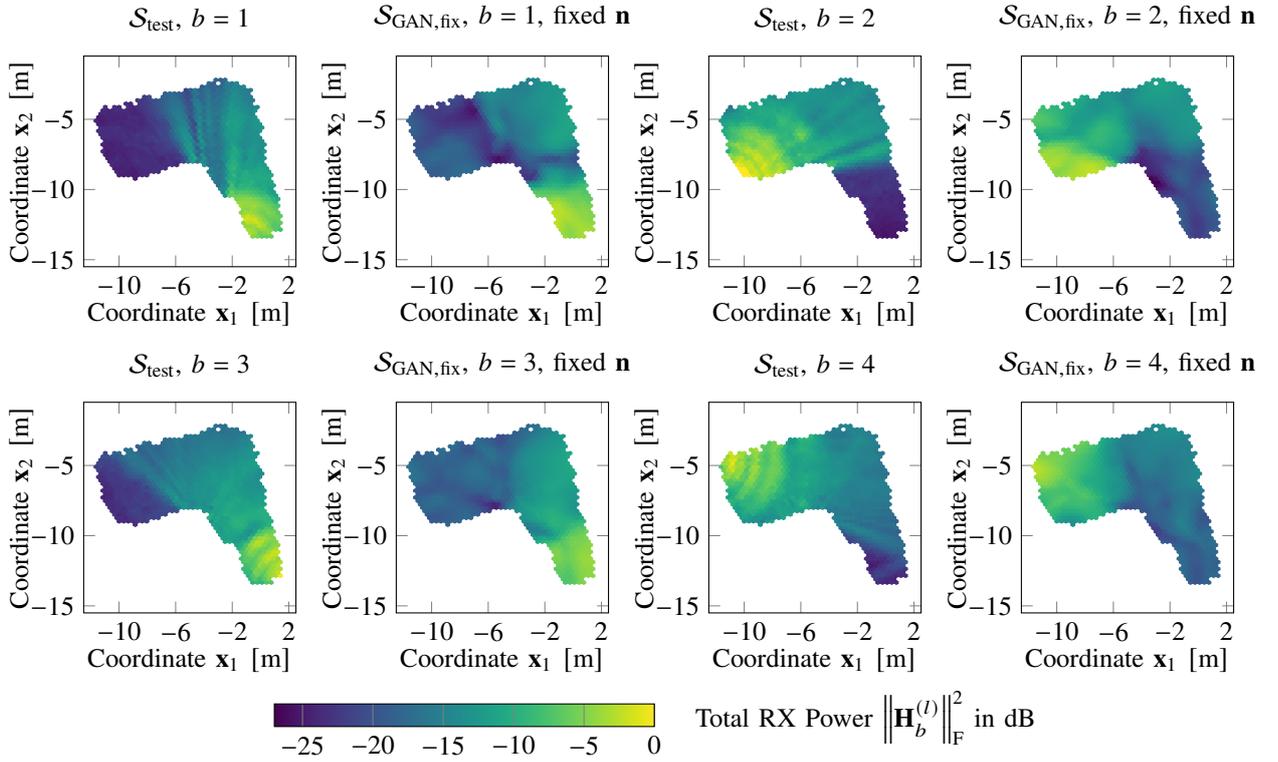
\begin{figure*}
    \centering
    \begin{subfigure}{0.22\textwidth}
        \begin{tikzpicture}
            \begin{axis}[
                width=0.7\columnwidth,
                height=0.7\columnwidth,
                scale only axis,
                xmin=-12.5,
                xmax=2.5,
                ymin=-15.5,
                ymax=-0.5,
                xlabel = {Coordinate $\mathbf x_1 ~ [\mathrm{m}]$},
                ylabel = {Coordinate $\mathbf x_2 ~ [\mathrm{m}]$},
                ylabel shift = -8 pt,
                xlabel shift = -4 pt,
                xtick={-10, -6, -2, 2},
                title={$\mathcal S_\mathrm{test}$, $b = 1$}
            ]
                \addplot[thick,blue] graphics[xmin=-12.5,ymin=-15.5,xmax=2.5,ymax=-0.5] {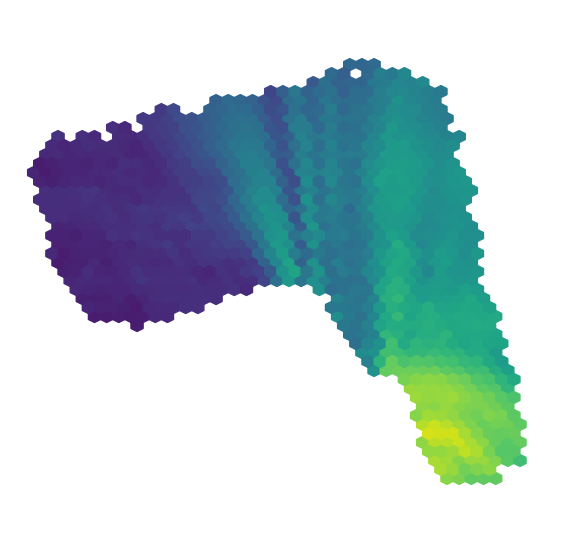};
            \end{axis}
        \end{tikzpicture}
    \end{subfigure}
    \begin{subfigure}{0.22\textwidth}
        \begin{tikzpicture}
            \begin{axis}[
                width=0.7\columnwidth,
                height=0.7\columnwidth,
                scale only axis,
                xmin=-12.5,
                xmax=2.5,
                ymin=-15.5,
                ymax=-0.5,
                xlabel = {Coordinate $\mathbf x_1 ~ [\mathrm{m}]$},
                ylabel = {Coordinate $\mathbf x_2 ~ [\mathrm{m}]$},
                ylabel shift = -8 pt,
                xlabel shift = -4 pt,
                xtick={-10, -6, -2, 2},
                title={$\mathcal S_\mathrm{GAN, fix}$, $b = 1$, fixed $\mathbf n$}
            ]
                \addplot[thick,blue] graphics[xmin=-12.5,ymin=-15.5,xmax=2.5,ymax=-0.5] {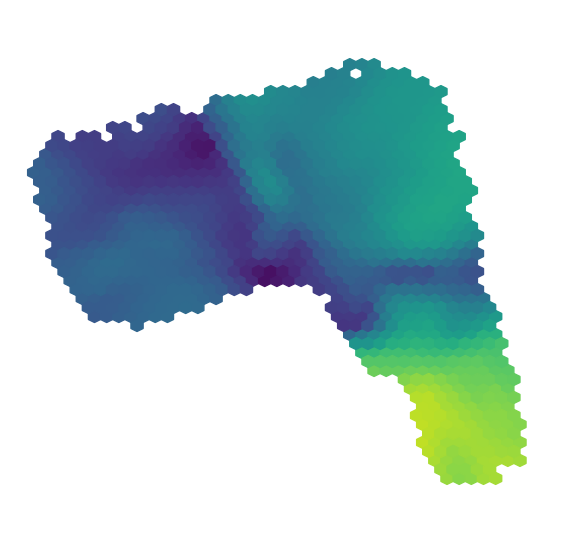};
            \end{axis}
        \end{tikzpicture}
    \end{subfigure}
    \begin{subfigure}{0.22\textwidth}
        \begin{tikzpicture}
            \begin{axis}[
                width=0.7\columnwidth,
                height=0.7\columnwidth,
                scale only axis,
                xmin=-12.5,
                xmax=2.5,
                ymin=-15.5,
                ymax=-0.5,
                xlabel = {Coordinate $\mathbf x_1 ~ [\mathrm{m}]$},
                ylabel = {Coordinate $\mathbf x_2 ~ [\mathrm{m}]$},
                ylabel shift = -8 pt,
                xlabel shift = -4 pt,
                xtick={-10, -6, -2, 2},
                title={$\mathcal S_\mathrm{test}$, $b = 2$}
            ]
                \addplot[thick,blue] graphics[xmin=-12.5,ymin=-15.5,xmax=2.5,ymax=-0.5] {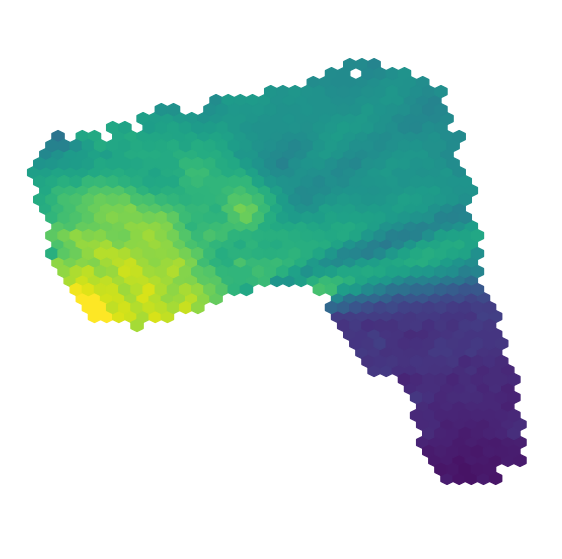};
            \end{axis}
        \end{tikzpicture}
    \end{subfigure}
    \begin{subfigure}{0.22\textwidth}
        \begin{tikzpicture}
            \begin{axis}[
                width=0.7\columnwidth,
                height=0.7\columnwidth,
                scale only axis,
                xmin=-12.5,
                xmax=2.5,
                ymin=-15.5,
                ymax=-0.5,
                xlabel = {Coordinate $\mathbf x_1 ~ [\mathrm{m}]$},
                ylabel = {Coordinate $\mathbf x_2 ~ [\mathrm{m}]$},
                ylabel shift = -8 pt,
                xlabel shift = -4 pt,
                xtick={-10, -6, -2, 2},
                title={$\mathcal S_\mathrm{GAN, fix}$, $b = 2$, fixed $\mathbf n$}
            ]
                \addplot[thick,blue] graphics[xmin=-12.5,ymin=-15.5,xmax=2.5,ymax=-0.5] {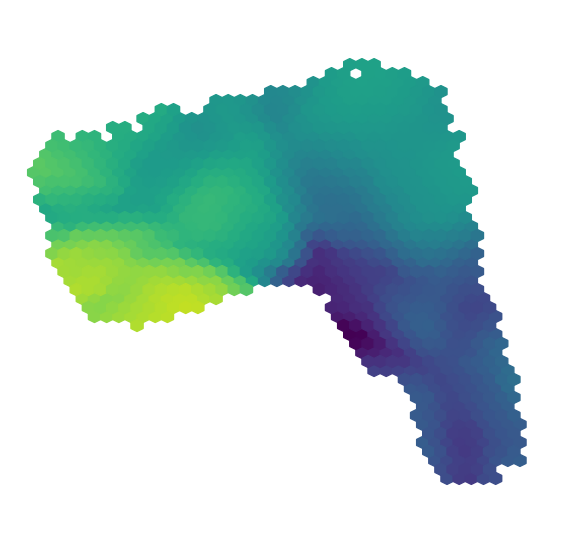};
            \end{axis}
        \end{tikzpicture}
    \end{subfigure}

    \vspace{0.1cm}

    \begin{subfigure}{0.22\textwidth}
        \begin{tikzpicture}
            \begin{axis}[
                width=0.7\columnwidth,
                height=0.7\columnwidth,
                scale only axis,
                xmin=-12.5,
                xmax=2.5,
                ymin=-15.5,
                ymax=-0.5,
                xlabel = {Coordinate $\mathbf x_1 ~ [\mathrm{m}]$},
                ylabel = {Coordinate $\mathbf x_2 ~ [\mathrm{m}]$},
                ylabel shift = -8 pt,
                xlabel shift = -4 pt,
                xtick={-10, -6, -2, 2},
                title={$\mathcal S_\mathrm{test}$, $b = 3$}
            ]
                \addplot[thick,blue] graphics[xmin=-12.5,ymin=-15.5,xmax=2.5,ymax=-0.5] {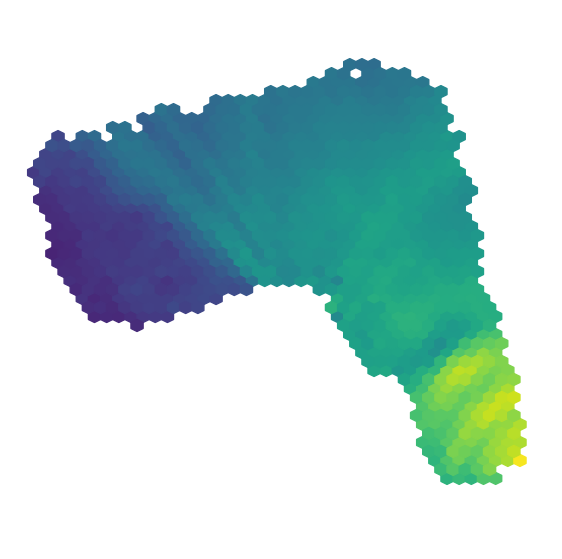};
            \end{axis}
        \end{tikzpicture}
    \end{subfigure}
    \begin{subfigure}{0.22\textwidth}
        \begin{tikzpicture}
            \begin{axis}[
                width=0.7\columnwidth,
                height=0.7\columnwidth,
                scale only axis,
                xmin=-12.5,
                xmax=2.5,
                ymin=-15.5,
                ymax=-0.5,
                xlabel = {Coordinate $\mathbf x_1 ~ [\mathrm{m}]$},
                ylabel = {Coordinate $\mathbf x_2 ~ [\mathrm{m}]$},
                ylabel shift = -8 pt,
                xlabel shift = -4 pt,
                xtick={-10, -6, -2, 2},
                title={$\mathcal S_\mathrm{GAN, fix}$, $b = 3$, fixed $\mathbf n$}
            ]
                \addplot[thick,blue] graphics[xmin=-12.5,ymin=-15.5,xmax=2.5,ymax=-0.5] {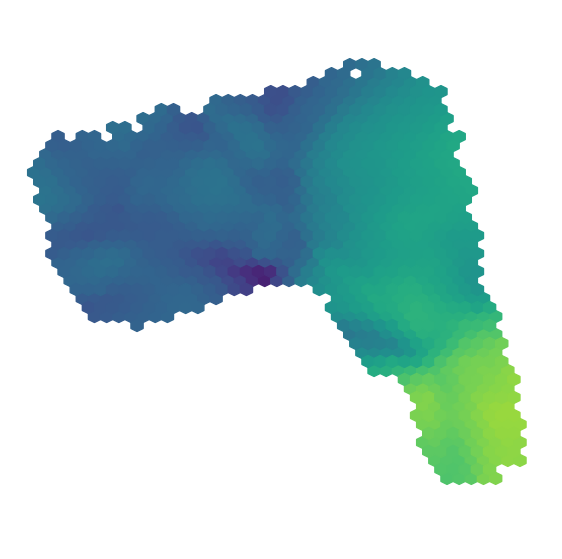};
            \end{axis}
        \end{tikzpicture}
    \end{subfigure}
    \begin{subfigure}{0.22\textwidth}
        \begin{tikzpicture}
            \begin{axis}[
                width=0.7\columnwidth,
                height=0.7\columnwidth,
                scale only axis,
                xmin=-12.5,
                xmax=2.5,
                ymin=-15.5,
                ymax=-0.5,
                xlabel = {Coordinate $\mathbf x_1 ~ [\mathrm{m}]$},
                ylabel = {Coordinate $\mathbf x_2 ~ [\mathrm{m}]$},
                ylabel shift = -8 pt,
                xlabel shift = -4 pt,
                xtick={-10, -6, -2, 2},
                title={$\mathcal S_\mathrm{test}$, $b = 4$}
            ]
                \addplot[thick,blue] graphics[xmin=-12.5,ymin=-15.5,xmax=2.5,ymax=-0.5] {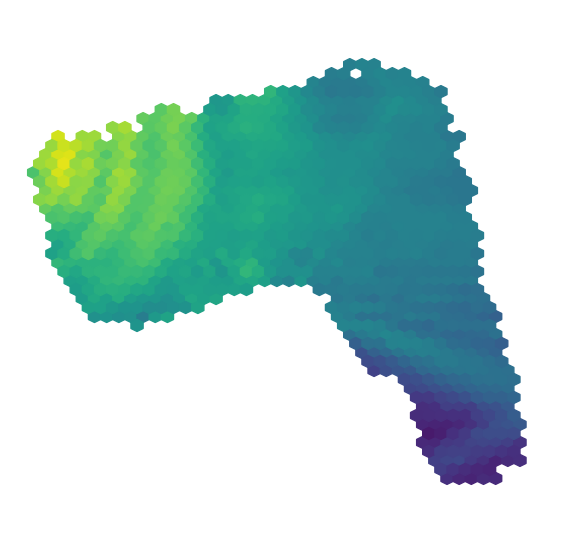};
            \end{axis}
        \end{tikzpicture}
    \end{subfigure}
    \begin{subfigure}{0.22\textwidth}
        \begin{tikzpicture}
            \begin{axis}[
                width=0.7\columnwidth,
                height=0.7\columnwidth,
                scale only axis,
                xmin=-12.5,
                xmax=2.5,
                ymin=-15.5,
                ymax=-0.5,
                xlabel = {Coordinate $\mathbf x_1 ~ [\mathrm{m}]$},
                ylabel = {Coordinate $\mathbf x_2 ~ [\mathrm{m}]$},
                ylabel shift = -8 pt,
                xlabel shift = -4 pt,
                xtick={-10, -6, -2, 2},
                title={$\mathcal S_\mathrm{GAN, fix}$, $b = 4$, fixed $\mathbf n$}
            ]
                \addplot[thick,blue] graphics[xmin=-12.5,ymin=-15.5,xmax=2.5,ymax=-0.5] {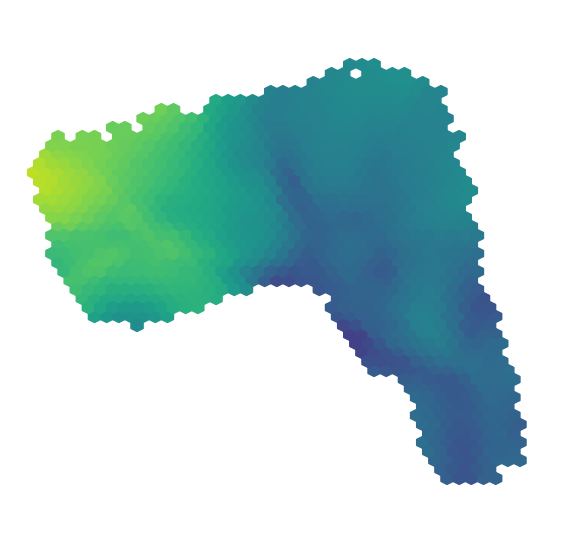};
            \end{axis}
        \end{tikzpicture}
    \end{subfigure}
    \begin{subfigure}[b]{0.5\textwidth}
        \centering
        \begin{tikzpicture}
            \begin{axis}[
                hide axis,
                scale only axis,
                colormap/viridis,
                colorbar,
                point meta min=-27,
                point meta max=0,
                colorbar horizontal,
                colorbar style={
                    width=5cm,
                    height=0.3cm,
                    xtick={-25,-20,-15,-10,-5,0},
                    xlabel = {Total RX Power $\left\lVert \mathbf H_b^{(l)} \right\rVert_\mathrm{F}^2$ in dB},
                    xlabel style = {
                        at = {(axis cs:15.0,2)}
                    }
                }]
                \addplot [draw=none] coordinates {(0,0)};
            \end{axis}
        \end{tikzpicture}
        \vspace{0.3cm}
    \end{subfigure}
    \vspace{-1cm}
    \caption{Spatial distribution of received powers $\lVert \mathbf H^{(l)}_{b} \rVert_\mathrm{F}^2$, for measured dataset $\mathcal S_\mathrm{meas}$ and generated dataset $\mathcal S_\mathrm{GAN, fixed}$. The power is normalized such that the maximum received power (over \emph{all} values $\lVert \mathbf H^{(l)}_{b} \rVert_\mathrm{F}^2$, measured and generated) is $0\,\mathrm{dB}$.}
    \label{fig:power-distribution}
\end{figure*}

Once \ac{GAN} training has finished, the critic is no longer needed, and the generator can be evaluated.
Depending on whether one views the generative channel model as an \emph{interpolator} or a \emph{random generator}, one may want to consider different evaluation criteria.
In this work, we compare both the spatial distributions of generated and measured \ac{CSI} as well as stochastic \ac{CSI} distributions.
In contrast to generative models in image processing, where subjective human perception of the quality of a generated image can play in important role in the evaluation process, humans cannot easily interpret \ac{CSI} data.
Instead, we need to rely on human-interpretable quantities that can be derived from channel tensors $\mathbf { \hat H }$ to judge the generated data.
For example, we can interpret and compare (to measured data) the distribution of powers, delay spreads or angular-domain properties in the generated data.

\subsection{Evaluated Channel Parameters and Statistics}
We focus our evaluation on three properties of the generated \ac{CSI}, those being the received power, the delay spread and the \ac{AoA}.
Furthermore, in an exemplary manner, we compare the delay spread distributions using the \ac{JSD}.

\subsubsection{Received Power}
The total received power over all antennas in one antenna array $b$ will be written as the norm
\[
    \left\lVert \mathbf H_{b} \right\rVert_\mathrm{F}^2 = \sum_{m_r = 1}^{M_r} \sum_{m_c = 1}^{M_c} \sum_{t = 1}^{N_\mathrm{tap}} \left|\mathbf H_{b, m_r, m_c, t}\right|^2.
\]

\subsubsection{Delay Spread}
\label{sec:delayspread}
For a single antenna in row $m_r$ and column $m_c$ of array $b$, the \ac{RMS} delay spread is defined as
\[
    \begin{split}
        \mathrm{DS}_{\mathrm{rms}, b, m_r, m_c} &= \sqrt{\frac{\sum_t (t - \bar t_{b, m_r, m_c})^2 \left|\mathbf H_{b, m_r, m_c, t}\right|^2}{\sum_t \left|\mathbf H_{b, m_r, m_c, t}\right|^2}} \\
        \text{with} ~~ \bar t_{b, m_r, m_c} &= \frac{\sum_t t \left|\mathbf H_{b, m_r, m_c, t}\right|^2}{\sum_t \left|\mathbf H_{b, m_r, m_c, t}\right|^2},
    \end{split}
\]
where $t = 1, \ldots, N_\mathrm{tap}$ is an index for the time tap.
To interpret the results, we may also want to visualize the \ac{RMS} delay spread averaged over all antennas in one array
\[
    \mathrm{\overline{DS}}_{\mathrm{rms}, b} = \frac{1}{M_r M_c} \sum_{m_r = 1}^{M_r} \sum_{m_c = 1}^{M_c} \mathrm{DS}_{\mathrm{rms}, b, m_r, m_c}.
\]
By comparing the measured and generated \ac{RMS} delay spread distribution, we can test if the \ac{GAN} learns meaningful channel \acp{PDP}.

\subsubsection{Angle of Arrival}
The array correlation matrix $\mathbf R^{(b)} \in \mathbb C^{M_\mathrm{c} \times M_\mathrm{c}}$ for the array with index $b$ is estimated over all rows of the \ac{UPA}.
That is, $\mathbf R^{(b)}$ may be used to estimate the azimuth \ac{AoA}, but contains no information about the elevation \ac{AoA}:
\[
    \mathbf { \hat R}^{(b)}_{m_{c1}, m_{c2}} = \sum_{m_\mathrm{r} = 1}^{M_\mathrm{r}} \sum_{t = 1}^{N_\mathrm{tap}} \mathbf H_{b, m_\mathrm{r}, m_\mathrm{c1}, t} \mathbf H_{b, m_\mathrm{r}, m_\mathrm{c2}, t}^* \in \mathbb C
\]
Under the assumption of a single source, we apply the root-\ac{MUSIC} algorithm to $\mathbf { \hat A}^{(b)}$ to find azimuth \ac{AoA} estimates $\hat \alpha^{(b)}$ from $\mathbf { \hat A}^{(b)}$ for each datapoint and each \ac{UPA}.
An azimuth \ac{AoA} of $\hat \alpha^{(b)} = 0^\circ$ indicates that the signal arrives directly from the front.
By comparing measured and generated \ac{AoA} distributions, we can test if the \ac{GAN} learns a meaningful distribution of channel coefficient phases.

\subsubsection{Jensen-Shannon Distance}
Furthermore, to quantify the similarity of different stochastic distributions, we use the \ac{JSD}.
For two distributions $P$ and $Q$ over the same sample space $\mathcal X$, the \ac{JSD} is defined as
\[
    \mathrm{d}_\mathrm{JS}(P || Q) = \sqrt{\frac{d_\mathrm{KL}(P || M) + d_\mathrm{KL}(Q || M)}{2}} ~~ \text{with} ~~ M = \frac{P + Q}{2},
\]
where $\mathrm d_\mathrm{KL}$ is the \ac{KL} divergence, given by
\[
    \mathrm d_\mathrm{KL}(P || Q) = \sum_{x \in \mathcal X} P(x) \log\left( \frac{P(x)}{Q(x)} \right).
\]

\subsection{GAN Evaluation Procedure}
\begin{figure}
    \centering
    \begin{subfigure}[b]{0.49\columnwidth}
        \centering
        \begin{tikzpicture}
            \begin{axis}[
                width=0.6\columnwidth,
                height=0.6\columnwidth,
                scale only axis,
                xmin=-12.5,
                xmax=2.5,
                ymin=-15.5,
                ymax=-0.5,
                xlabel = {Coordinate $\mathbf x_1 ~ [\mathrm{m}]$},
                ylabel = {Coordinate $\mathbf x_2 ~ [\mathrm{m}]$},
                ylabel shift = -8 pt,
                xlabel shift = -4 pt,
                xtick={-10, -6, -2, 2},
                xtick style={color=black}
            ]
                \addplot[thick,blue] graphics[xmin=-12.5,ymin=-15.5,xmax=2.5,ymax=-0.5] {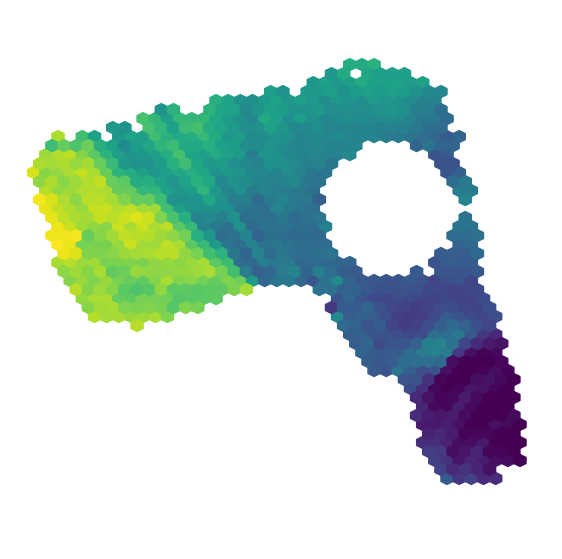};
            \end{axis}
        \end{tikzpicture}
        \vspace{-0.2cm}
        \caption{Training Set $\mathcal S_\mathrm{train}$}
        \label{fig:ds-trainingset}
    \end{subfigure}
    \begin{subfigure}[b]{0.49\columnwidth}
        \centering
        \begin{tikzpicture}
            \begin{axis}[
                width=0.6\columnwidth,
                height=0.6\columnwidth,
                scale only axis,
                xmin=-12.5,
                xmax=2.5,
                ymin=-15.5,
                ymax=-0.5,
                xlabel = {Coordinate $\mathbf x_1 ~ [\mathrm{m}]$},
                ylabel = {Coordinate $\mathbf x_2 ~ [\mathrm{m}]$},
                ylabel shift = -8 pt,
                xlabel shift = -4 pt,
                xtick={-10, -6, -2, 2},
                xtick style={color=black}
            ]
                \addplot[thick,blue] graphics[xmin=-12.5,ymin=-15.5,xmax=2.5,ymax=-0.5] {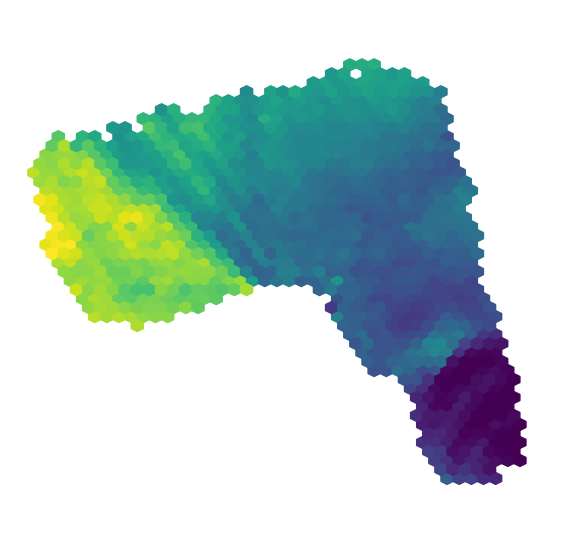};
            \end{axis}
        \end{tikzpicture}
        \vspace{-0.2cm}
        \caption{Test Set $\mathcal S_\mathrm{test}$}
        \label{fig:ds-testset}
    \end{subfigure}
    \begin{subfigure}[b]{0.49\columnwidth}
        \centering
        \begin{tikzpicture}
            \begin{axis}[
                width=0.6\columnwidth,
                height=0.6\columnwidth,
                scale only axis,
                xmin=-12.5,
                xmax=2.5,
                ymin=-15.5,
                ymax=-0.5,
                xlabel = {Coordinate $\mathbf x_1 ~ [\mathrm{m}]$},
                ylabel = {Coordinate $\mathbf x_2 ~ [\mathrm{m}]$},
                ylabel shift = -8 pt,
                xlabel shift = -4 pt,
                xtick={-10, -6, -2, 2}
            ]
                \addplot[thick,blue] graphics[xmin=-12.5,ymin=-15.5,xmax=2.5,ymax=-0.5] {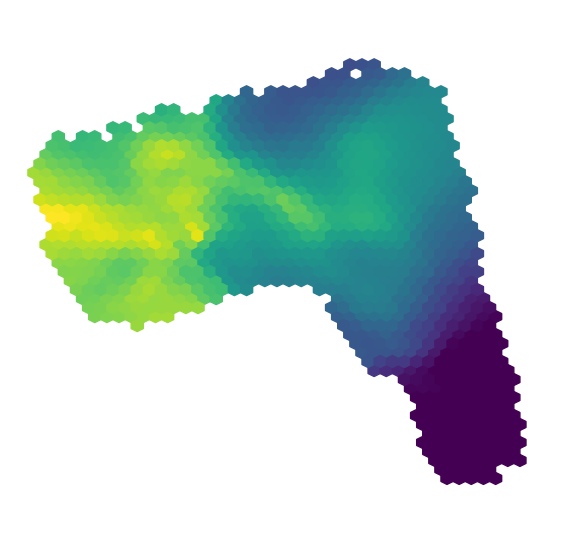};
            \end{axis}
        \end{tikzpicture}
        \vspace{-0.2cm}
        \caption{$\mathcal S_\mathrm{GAN, fix}$, fixed $\mathbf n^{(1)}$}
        \label{fig:ds-gan-fixed1}
    \end{subfigure}
    \begin{subfigure}[b]{0.49\columnwidth}
        \centering
        \begin{tikzpicture}
            \begin{axis}[
                width=0.6\columnwidth,
                height=0.6\columnwidth,
                scale only axis,
                xmin=-12.5,
                xmax=2.5,
                ymin=-15.5,
                ymax=-0.5,
                xlabel = {Coordinate $\mathbf x_1 ~ [\mathrm{m}]$},
                ylabel = {Coordinate $\mathbf x_2 ~ [\mathrm{m}]$},
                ylabel shift = -8 pt,
                xlabel shift = -4 pt,
                xtick={-10, -6, -2, 2}
            ]
                \addplot[thick,blue] graphics[xmin=-12.5,ymin=-15.5,xmax=2.5,ymax=-0.5] {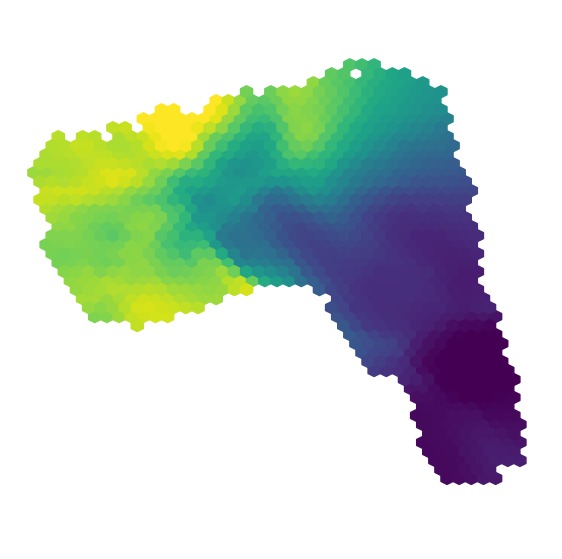};
            \end{axis}
        \end{tikzpicture}
        \vspace{-0.2cm}
        \caption{$\mathcal S_\mathrm{GAN, fix}$, fixed $\mathbf n^{(2)}$}
        \label{fig:ds-gan-fixed2}
    \end{subfigure}
    \begin{subfigure}[b]{0.49\columnwidth}
        \centering
        \begin{tikzpicture}
            \begin{axis}[
                width=0.6\columnwidth,
                height=0.6\columnwidth,
                scale only axis,
                xmin=-12.5,
                xmax=2.5,
                ymin=-15.5,
                ymax=-0.5,
                xlabel = {Coordinate $\mathbf x_1 ~ [\mathrm{m}]$},
                ylabel = {Coordinate $\mathbf x_2 ~ [\mathrm{m}]$},
            ylabel shift = -8 pt,
            xlabel shift = -4 pt,
            xtick={-10, -6, -2, 2}
        ]
            \addplot[thick,blue] graphics[xmin=-12.5,ymin=-15.5,xmax=2.5,ymax=-0.5] {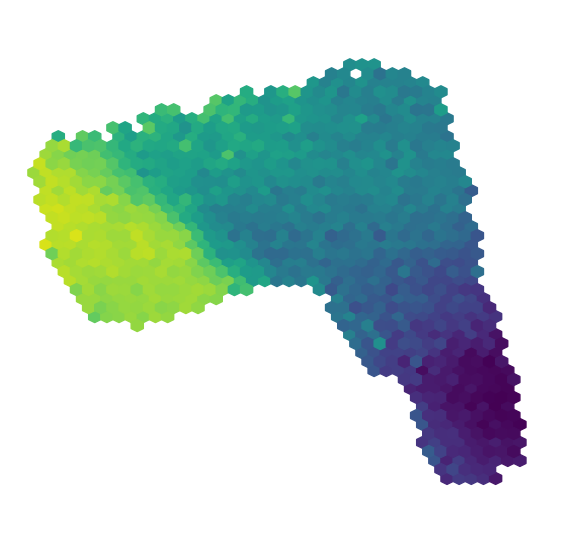};
        \end{axis}
        \end{tikzpicture}
        \vspace{-0.2cm}
        \caption{$\mathcal S_\mathrm{GAN, var}$, $\mathbf n \sim \mathcal N(0, 1)$}
        \label{fig:ds-gan-random}
    \end{subfigure}
    \begin{subfigure}[b]{0.49\columnwidth}
        \centering
        \begin{tikzpicture}
            \begin{axis}[
                width=0.6\columnwidth,
                height=0.6\columnwidth,
                scale only axis,
                xmin=-12.5,
                xmax=2.5,
                ymin=-15.5,
                ymax=-0.5,
                xlabel = {Coordinate $\mathbf x_1 ~ [\mathrm{m}]$},
                ylabel = {Coordinate $\mathbf x_2 ~ [\mathrm{m}]$},
                ylabel shift = -8 pt,
                xlabel shift = -4 pt,
                xtick={-10, -6, -2, 2}
            ]
                \addplot[thick,blue] graphics[xmin=-12.5,ymin=-15.5,xmax=2.5,ymax=-0.5] {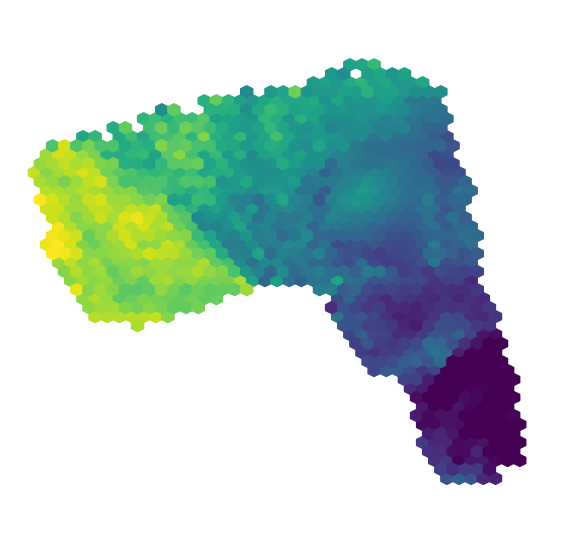};
            \end{axis}
        \end{tikzpicture}
        \vspace{-0.2cm}
        \caption{Baseline $\mathcal S_\mathrm{interp}$}
        \label{fig:ds-linear-interp}
    \end{subfigure}
    \begin{subfigure}[b]{0.5\textwidth}
        \centering
        \begin{tikzpicture}
            \begin{axis}[
                hide axis,
                scale only axis,
                colormap/viridis,
                colorbar,
                point meta min=30,
                point meta max=130,
                colorbar horizontal,
                colorbar style={
                    width=4cm,
                    height=0.3cm,
                    xtick={30, 50, 100, 130},
                    xlabel = {Mean RMS Delay Spread $\mathrm{\overline{DS}}_{\mathrm{rms}, 1}$ [ns]},
                    xlabel style = {
                        at = {(axis cs:80.0,3.5)}
                    }
                }]
                \addplot [draw=none] coordinates {(0,0)};
            \end{axis}
        \end{tikzpicture}
        \vspace{-0.2cm}
    \end{subfigure}
    \caption{RMS delay spread $\mathrm{\overline{DS}}_{\mathrm{rms}, b}$, averaged over all antennas in array $b = 1$.}
    \label{fig:ds-fixed-n}
    \vspace{-0.3cm}
\end{figure}
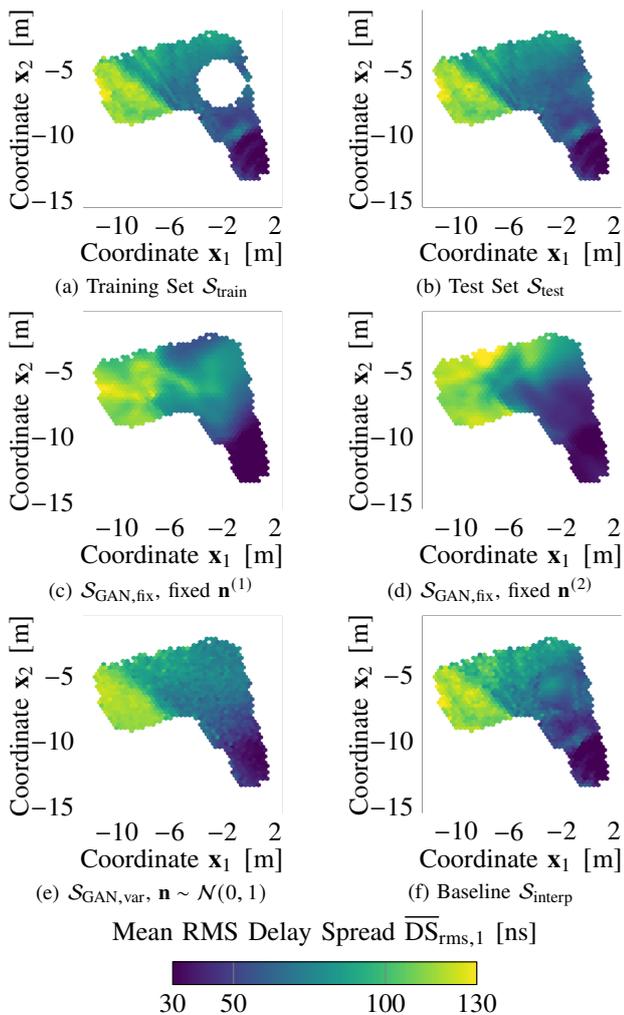

To obtain a generated dataset $\mathcal S_\mathrm{gen}$ that is comparable to the test set $\mathcal S_\mathrm{test}$, we use $G$ to produce channel realizations for all positions $\mathbf x^{(l)}$ present in $\mathcal S_\mathrm{test}$.
Evaluating the generator can be approached in two ways:
\begin{enumerate}[label=(\Roman*)]
    \item\label{approach:fixed} The noise vector $\mathbf n \sim \mathcal N(0, 1)$ may be drawn once and then remain fixed over all positions (conditions) $\mathbf x$:
    \begin{align*}
        \mathcal S_\mathrm{GAN, fix} = \left\{ \left(G\left(\mathbf x^{(l)}, \mathbf n \right), \mathbf x^{(l)}\right) \right\}_{(\mathbf H^{(l)}, \mathbf x^{(l)}) \in \mathcal S_\mathrm{test}}, \\ ~ \text{fixed} ~ \mathbf n \sim \mathcal N(0, 1)
    \end{align*}
    \item\label{approach:variable} The noise vector $\mathbf n^{(l)} \sim \mathcal N(0, 1)$ may be drawn randomly for each position (condition) $\mathbf x^{(l)}$:
    \[
        \mathcal S_\mathrm{GAN, var} = \left\{ \left(G\left(\mathbf x^{(l)}, \mathbf n^{(l)}\right), \mathbf x^{(l)}\right) \right\}_{(\mathbf H^{(l)}, \mathbf x^{(l)}) \in \mathcal S_\mathrm{test}, ~ \mathbf n^{(l)} \sim \mathcal N(0, 1)}
    \]
\end{enumerate}
The first approach should lead to a higher level of spatial consistency in the generated \ac{CSI} data, whereas the second approach can reveal what the generator has learned about the spatial channel distribution.

\subsection{Linear Interpolator Evaluation Procedure}
To generate \ac{CSI} with the linear interpolation baseline, which makes use of the entire training set $\mathcal S_\mathrm{train}$ to parametrize the interpolant, we evaluate the interpolant $F_\mathrm{interp}$ at the test set locations, yielding the interpolated \ac{CSI} dataset:
\[
    \mathcal S_\mathrm{interp} = \left\{ \left(F_\mathrm{interp}\left(\mathbf x^{(l)}\right), \mathbf x^{(l)}\right) \right\}_{(\mathbf H^{(l)}, \mathbf x^{(l)}) \in \mathcal S_\mathrm{test}}
\]

\subsection{Spatial Distribution Analysis}
\label{sec:spatialdistribution}
We consider the distribution of generated \ac{CSI} relative to the two-dimensional spatial coordinates, which are passed to the \ac{GAN} as condition.
If the \ac{CSI} generated by the \ac{GAN} is similar to the test set \ac{CSI}, this could indicate that \ac{GAN}-based channel models are useful as \emph{inter-/extrapolators}.
First, we evaluate the generator output distribution according to approach \ref{approach:fixed}, i.e., for fixed input noise $\mathbf n$.
The corresponding received power distributions $\lVert \mathbf H^{(l)}_{b} \rVert_\mathrm{F}^2$ for each antenna array $b = 1, \ldots, 4$, both measured and generated for one particular noise realization, are shown in Fig.~\ref{fig:power-distribution}.
Clearly, the \ac{WGAN} has learned to produce \ac{CSI} with some spatial consistency, and the spatial distribution of generated and measured channels is similar, but not identical.
The generated channel realizations greatly depend on the choice of receiver antenna array $b$.
However, the neural network fails to produce \ac{CSI} with true spatial consistency.
For example, the concentric circles around the antenna arrays visible in the measured data (which are likely caused by a strong floor reflection) are not visible in the generated power distribution.

We can also analyze the \ac{RMS} delay spread for approach~\ref{approach:fixed}.
The results for array $b = 1$ are shown in Fig.~\ref{fig:ds-gan-fixed1} and Fig.~\ref{fig:ds-gan-fixed2}, for two different fixed noise vectors $\mathbf n^{(1)}$ and $\mathbf n^{(2)}$:
It can be seen that the delay spread distributions depend on the noise vector and that they are also spatially consistent, yet distinct from the measured distribution.
However, from Fig.~\ref{fig:ds-linear-interp}, it is also clear that the linear interpolation baseline produces \ac{CSI} that much more closely resembles the test set if accurate \ac{CSI} prediction is desired.
Though not shown in Fig.~\ref{fig:ds-fixed-n}, these observations also hold for antenna arrays $b = 2, 3, 4$.
If we choose a random noise realization for every position $\mathbf x$ as in approach \ref{approach:variable}, the spatial \ac{RMS} delay spread distribution more closely resembles the test set, as visible in Fig.~\ref{fig:ds-gan-random}.

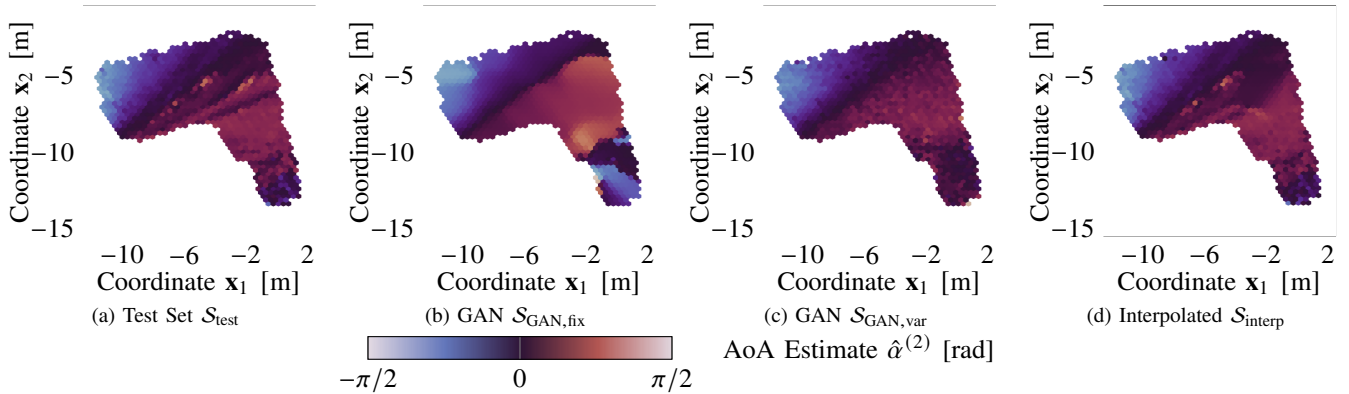
\begin{figure*}
    \centering
    \begin{subfigure}[b]{0.24\textwidth}
        \centering
        \begin{tikzpicture}
            \begin{axis}[
                width=0.7\columnwidth,
                height=0.7\columnwidth,
                scale only axis,
                xmin=-12.5,
                xmax=2.5,
                ymin=-15.5,
                ymax=-0.5,
                xlabel = {Coordinate $\mathbf x_1 ~ [\mathrm{m}]$},
                ylabel = {Coordinate $\mathbf x_2 ~ [\mathrm{m}]$},
                ylabel shift = -8 pt,
                xlabel shift = -4 pt,
                xtick={-10, -6, -2, 2}
            ]
                \addplot[thick,blue] graphics[xmin=-12.5,ymin=-15.5,xmax=2.5,ymax=-0.5] {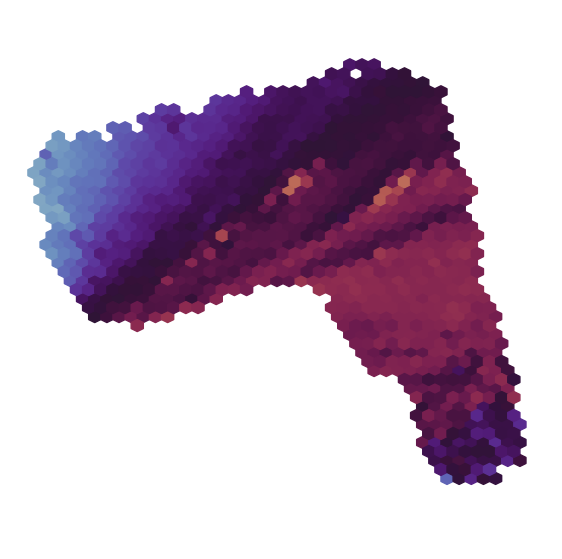};
            \end{axis}
        \end{tikzpicture}
        \vspace{-0.2cm}
        \caption{Test Set $\mathcal S_\mathrm{test}$}
        \label{fig:aoa-testset}
    \end{subfigure}
    \begin{subfigure}[b]{0.24\textwidth}
        \centering
        \begin{tikzpicture}
            \begin{axis}[
                width=0.7\columnwidth,
                height=0.7\columnwidth,
                scale only axis,
                xmin=-12.5,
                xmax=2.5,
                ymin=-15.5,
                ymax=-0.5,
                xlabel = {Coordinate $\mathbf x_1 ~ [\mathrm{m}]$},
                ylabel = {Coordinate $\mathbf x_2 ~ [\mathrm{m}]$},
                ylabel shift = -8 pt,
                xlabel shift = -4 pt,
                xtick={-10, -6, -2, 2}
            ]
                \addplot[thick,blue] graphics[xmin=-12.5,ymin=-15.5,xmax=2.5,ymax=-0.5] {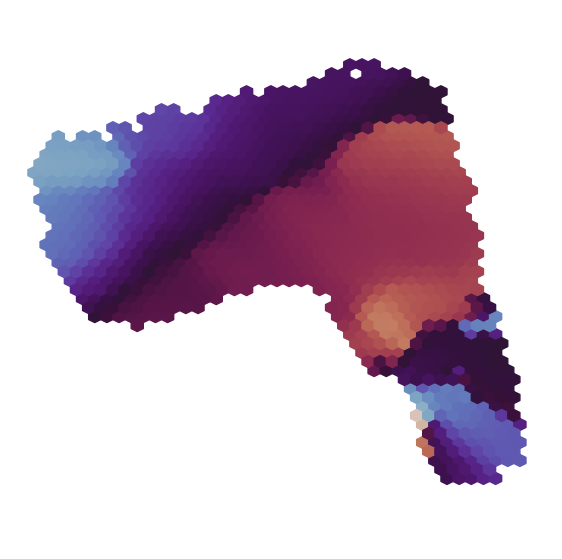};
            \end{axis}
        \end{tikzpicture}
        \vspace{-0.2cm}
        \caption{GAN $\mathcal S_\mathrm{GAN, fix}$}
        \label{fig:aoa-gan-fixed}
    \end{subfigure}
    \begin{subfigure}[b]{0.24\textwidth}
        \centering
        \begin{tikzpicture}
            \begin{axis}[
                width=0.7\columnwidth,
                height=0.7\columnwidth,
                scale only axis,
                xmin=-12.5,
                xmax=2.5,
                ymin=-15.5,
                ymax=-0.5,
                xlabel = {Coordinate $\mathbf x_1 ~ [\mathrm{m}]$},
                ylabel = {Coordinate $\mathbf x_2 ~ [\mathrm{m}]$},
                ylabel shift = -8 pt,
                xlabel shift = -4 pt,
                xtick={-10, -6, -2, 2}
            ]
                \addplot[thick,blue] graphics[xmin=-12.5,ymin=-15.5,xmax=2.5,ymax=-0.5] {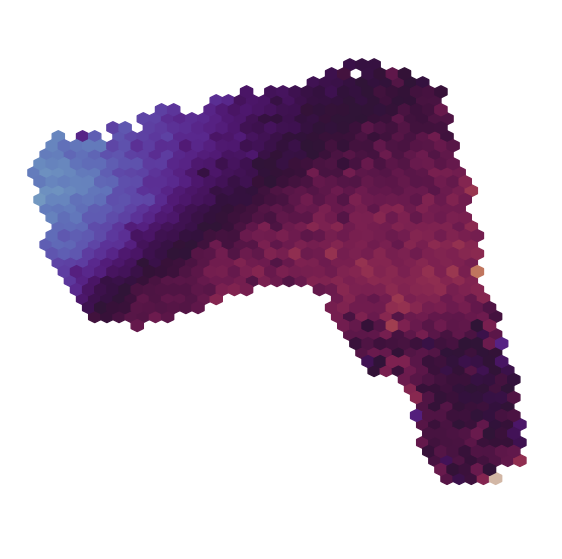};
            \end{axis}
        \end{tikzpicture}
        \vspace{-0.2cm}
        \caption{GAN $\mathcal S_\mathrm{GAN, var}$}
        \label{fig:aoa-gan-random}
    \end{subfigure}
    \begin{subfigure}[b]{0.24\textwidth}
        \centering
        \begin{tikzpicture}
            \begin{axis}[
                width=0.7\columnwidth,
                height=0.7\columnwidth,
                scale only axis,
                xmin=-12.5,
                xmax=2.5,
                ymin=-15.5,
                ymax=-0.5,
                xlabel = {Coordinate $\mathbf x_1 ~ [\mathrm{m}]$},
                ylabel = {Coordinate $\mathbf x_2 ~ [\mathrm{m}]$},
                ylabel shift = -8 pt,
                xlabel shift = -4 pt,
                xtick={-10, -6, -2, 2}
            ]
                \addplot[thick,blue] graphics[xmin=-12.5,ymin=-15.5,xmax=2.5,ymax=-0.5] {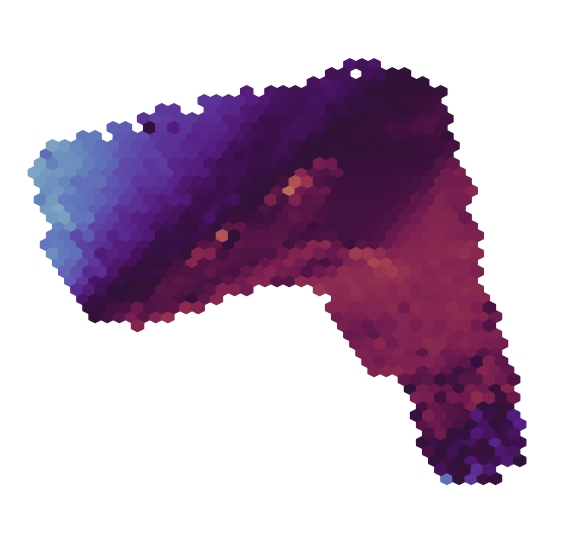};
            \end{axis}
        \end{tikzpicture}
        \vspace{-0.2cm}
        \caption{Interpolated $\mathcal S_\mathrm{interp}$}
        \label{fig:aoa-linear-interp}
    \end{subfigure}
    \begin{subfigure}[b]{0.5\textwidth}
        \centering
        \vspace{-0.2cm}
        \begin{tikzpicture}
            \begin{axis}[
                hide axis,
                scale only axis,
                colormap/twilight,
                colorbar,
                point meta min=-1.57,
                point meta max=1.57,
                colorbar horizontal,
                colorbar style={
                    width=4cm,
                    height=0.3cm,
                    xtick={-1.57, 0, 1.57},
                    xticklabels={$-\pi / 2$, $0$, $\pi / 2$},
                    xlabel = {AoA Estimate $\hat \alpha^{(2)}$ [rad]},
                    xlabel style = {
                        at = {(axis cs:3.5,1.5)}
                    }
                }]
                \addplot [draw=none] coordinates {(0,0)};
            \end{axis}
        \end{tikzpicture}
        \vspace{-0.3cm}
    \end{subfigure}
    \vspace{-0.1cm}
    \caption{Spatial azimuth AoA estimate distribution for antenna array $b = 2$ for various measured / generated \ac{CSI} datasets. Note that the color map is circular.}
    \label{fig:aoa-distribution}
    \vspace{-0.3cm}
\end{figure*}

The observed spatial distribution of received powers and delay spreads indicates that the generator was at least able to learn meaningful \acp{PDP} of the wireless channel.
To test if the generator has also learned a meaningful distribution of phase information, we visualize the spatial distribution of \ac{AoA} estimates generated by \ac{MUSIC} for measured and generated \ac{CSI} data in Fig.~\ref{fig:aoa-distribution}.
Clearly, the \ac{GAN} has captured the relationship between location and \ac{CSI} phases to some extent, as indicated by the clearly visible gradient in the \ac{AoA} estimates in Fig.~\ref{fig:aoa-gan-random}.
The \ac{AoA} estimate for the generated \ac{CSI} corresponds to the true \ac{AoA} of the \ac{LoS} regions in the dataset.
However, the linear interpolation baseline also clearly outperforms the \ac{GAN} if the objective is to generate true-to-life \ac{CSI} data: The \ac{AoA} estimates for the interpolated \ac{CSI} in Fig.~\ref{fig:aoa-linear-interp} are almost identical to the \ac{AoA} estimates generated for the test set in Fig.~\ref{fig:aoa-testset}.
Fig.~\ref{fig:aoa-gan-fixed}, which shows the \ac{AoA} estimate distribution for \ac{CSI} generated by a \ac{GAN} with fixed noise realization looks rather artificial:
The individual \ac{AoA} estimates for the generated \ac{CSI} are all plausible for the specific position $\mathbf x$, but the spatial distribution is unrealistic, with constant \ac{AoA} in some areas and rapid changes over space elsewhere.
This may indicate that the generator has failed to produce \ac{CSI} with meaningful spatial consistency.

Comparing test set distribution and generated distributions of received power, \ac{RMS} delay spread and \ac{AoA} estimates, we conclude that a \ac{GAN}-based channel model is not a good spatial interpolator in our scenario, as even a simple linear interpolation baseline outperforms the \ac{GAN}:
Judging by \ac{RMS} delay spread and \ac{AoA} estimates, the interpolated \ac{CSI} is clearly more similar to the test set than the GAN-generated \ac{CSI}.

\subsection{Stochastic Distribution Analysis}
\label{sec:stochasticdistribution}

\begin{table}
    \centering
    \begin{tabular}{r | c c c c c c}
        $d_\mathrm{JS}$ & $\mathcal S_\mathrm{train}$ & $\mathcal S_\mathrm{test}$ & $\mathcal S_\mathrm{GAN,var}$ & $\mathcal S_\mathrm{GAN,fix}$ & $\mathcal S_\mathrm{interp}$ & Gauss. \\ \hline
        $\mathcal S_\mathrm{train}$ & 0.000 & 0.041 & 0.066 & 0.093 & 0.049 & 0.160 \\
        $\mathcal S_\mathrm{test}$ & 0.041 & 0.000 & 0.082 & 0.115 & 0.067 & 0.149 \\
        $\mathcal S_\mathrm{GAN,var}$ & 0.066 & 0.082 & 0.000 & 0.053 & 0.054 & 0.162 \\
        $\mathcal S_\mathrm{GAN,fix}$ & 0.093 & 0.115 & 0.053 & 0.000 & 0.070 & 0.191 \\
        $\mathcal S_\mathrm{interp}$ & 0.049 & 0.067 & 0.054 & 0.070 & 0.000 & 0.156 \\
        Gauss. & 0.160 & 0.149 & 0.162 & 0.191 & 0.156 & 0.000
    \end{tabular}
    \caption{JSD between various RMS delay spread distributions.}
    \label{tab:jsd}
    \vspace{-0.4cm}
\end{table}

In the previous section, we compared the spatial distribution of generated \ac{CSI} to the test set.
If one is of the opinion that the \ac{GAN}-based channel model should be used as a \emph{random generator} for \ac{CSI} that serves to generate \ac{CSI} data with a certain stochastic distribution (and not to interpolate \ac{CSI}), this is not a fair comparison.
Therefore, we now compare the stochastic distribution of measured and generated \ac{CSI}.

For the following analysis, we focus on the \ac{RMS} delay spread only and disregard all spatial aspects.
Of course, these analyses could also be performed on other \ac{CSI} properties like received power or angular-domain information, with similar conclusions.
We estimate the density function of the \ac{RMS} delay spreads over all antennas in all antenna arrays for measured and generated \ac{CSI} by binning delay spreads into 150 bins.
In addition, we compare the generated \ac{RMS} delay spread distribution to randomly drawing from a Gaussian \ac{PDF}, whose first- and second-order moments are directly estimated from the training set's \ac{RMS} delay spread distribution.

\begin{figure}
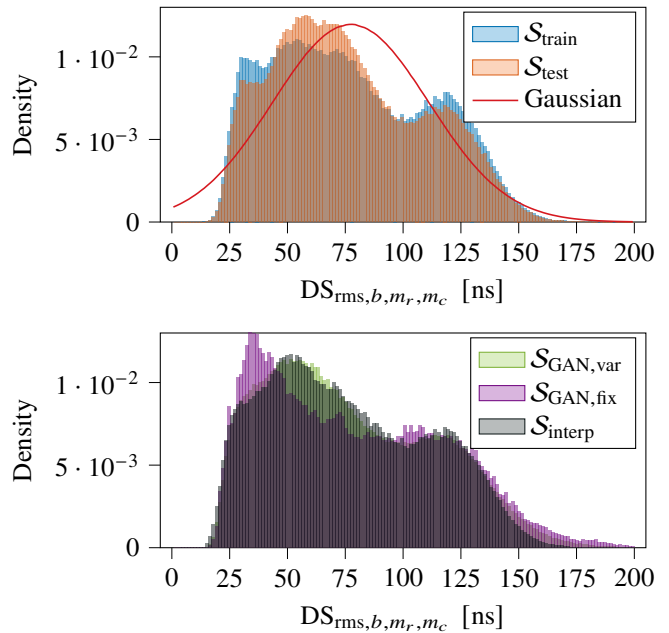

    \centering
    \begin{subfigure}{\columnwidth}
        \traintrue
        \testtrue
        \ganvarfalse
        \ganfixfalse
        \interpfalse
        \gaussiantrue
        \include{fig/delay-spread-histogram}
        \vspace{-0.6cm}
    \end{subfigure}
    \begin{subfigure}{\columnwidth}
        \trainfalse
        \testfalse
        \ganvartrue
        \ganfixtrue
        \interptrue
        \gaussianfalse
        \include{fig/delay-spread-histogram}
        \vspace{-0.3cm}
    \end{subfigure}
    \vspace{-1.3cm}
    \caption{Density functions of RMS delay spread distributions for various CSI datasets and Gaussian approximation.}
    \label{fig:ds-distribution}
    \vspace{-0.3cm}
\end{figure}

The resulting histograms and the parametrized Gaussian distribution are shown in Fig.~\ref{fig:ds-distribution}.
The high degree of similarity between the measured and generated histograms indicates that the \ac{GAN} is able to approximate the measured channel distribution well, but so is the linear interpolator.
Furthermore, the histogram indicates that the \ac{GAN} not only learned the first and second-order moments of the delay spread distribution, but also higher-order moments.
If one were instead to sample \ac{RMS} delay spreads for a generative stochastic model from an estimated Gaussian distribution, the resulting \ac{CSI} distribution would be a bad approximation of the realistic \ac{CSI}.
Please note that the \ac{RMS} delay spread distribution for $\mathcal S_\mathrm{GAN, fix}$ depends on the particular random noise input $\mathbf n$, so the density shown here is only exemplary.

To quantify the similarity of the densities in Fig.~\ref{fig:ds-distribution} more objectively, we compute the \acp{JSD} between any two density functions, resulting in the \ac{JSD} distance matrix shown in Tab.~\ref{tab:jsd}.
Unsurprisingly, the \ac{RMS} delay spread distribution that most closely resembles the test set is the training set, followed by the interpolated \ac{CSI}.
The \ac{RMS} delay spread distribution generated by the \ac{GAN} is less similar than the interpolated \ac{CSI}, but still much better than sampling from a Gaussian distribution.

We conclude that in our scenario, the \ac{GAN}-based channel model was able to reproduce the statistical \ac{CSI} distribution, at least when judged on the basis of the delay spread of generated \ac{CSI}.
However, the \ac{CSI} that the \ac{GAN} generates is highly specific to the considered setup and environment.

\section{Criticism}
\newif\iftrain
\newif\iftest
\newif\ifganvar
\newif\ifganfix
\newif\ifinterp
\newif\ifgaussian

Neural networks show promising results for various communications-related applications, including channel coding \cite{gruber2017deep}, channel charting \cite{studer2018channel}, and channel prediction \cite{jiang2019neural}.
What these applications have in common is that there is a lot of training data available, but there is no good system model due to the unknown statistical properties of the propagation channel.
Clearly, under these circumstances, machine learning can lead to significant improvements.
On the face of it, the same logic applies to \ac{GAN}-based channel modeling, where large \ac{CSI} datasets are available for training.
However, upon closer inspection, this impression is deceptive: Just because there is a lot of training data available does not necessarily mean that the training data is suitable for training a useful generative model.
In the following, we outline why we believe \ac{GAN}-based channel modeling (and generative neural network-based channel modeling in general) is not ready for practical use in its current form.
In Sec. \ref{sec:spatialdistribution} we already concluded that \ac{GAN}-based channel models are not good interpolators.
Here, we argue that in their current form, they are also not good \ac{GSCM} replacements and that significant architectural changes are necessary for them to become useful.

\subsection{Lack of generalizability}
A \ac{GAN}-based channel model is trained on one particular dataset or a collection of \ac{CSI} datasets captured with the same setup.
The current system architecture would not allow for training the same generator on datasets captured in different environments with different antenna setups.
Therefore, the \ac{GAN}-based channel model is always specific to the environment it was trained for and, in contrast to a \ac{GSCM}, cannot generalize to a different number of antennas, antenna deployments, carrier frequencies or other environment parameters.

\subsection{Lack of meaningful spatial consistency}
Spatial consistency is an important concept in channel modeling:
In \acp{GSCM}, it can be ensured in various ways, e.g., by defining scattering clusters \cite{burkhardt2014quadriga}, or by modeling spatial correlations of stochastic distributions like, for example, spatially correlated shadow fading \cite{cai2003two}.
Judging by the results in Sec. \ref{sec:spatialdistribution}, it may seem like the \ac{GAN} ensures at least some level of spatial consistency.
This is the case in the sense that the \ac{GAN} generates a \ac{CSI} sample that is plausible for the considered position (condition) $\mathbf x$.
However, it does not mean that the \ac{GAN} ensures any meaningful spatial correlation between two generated \ac{CSI} samples $\hat {\mathbf H}^{(1)}$ and $\hat {\mathbf H}^{(2)}$ at different locations $\mathbf x^{(1)}$ and $\mathbf x^{(2)}$.
In fact, we found that the level of spatial correlation depends on the range to which the condition $\mathbf x$ is normalized.
This indicates that spatial correlation is not inferred from the training data, but is merely an artifact of the network architecture and training procedure.

\subsection{Lack of interpretability}
One of the strengths of channel models is the ability to see how a designed system would perform under different conditions.
For example, a system designer may want to understand how the system performs at different carrier frequencies, for different antenna deployments, Doppler spreads, path loss exponents or noise distributions.
Other than in \acp{GSCM}, these properties are not controlled in \ac{GAN}-based channel models.
If a developed system or algorithm shows good or poor performance on a specific \ac{GAN}-based channel model, it would be difficult to understand why that is the case.

\subsection{Availability of a dataset}
The results in Sec. \ref{sec:stochasticdistribution} indicate that using a \ac{GAN}-based channel model is only applicable if the objective is to have a channel model for the particular environment that the measured (or simulated) training dataset was captured in.
In that case, a \ac{GAN} can generate \ac{CSI} that follows the distribution of the training dataset.
However, if the objective is to produce \ac{CSI} that is as similar as possible to a training set, one may just as well use the training dataset itself.
If the dataset is too small, one could use various classical data augmentation techniques (e.g., adding noise, adding random global phase rotations) or interpolation, rendering the generative model unnecessary.

\section{Conclusion}
We have demonstrated that a \ac{GAN} can be used to learn the statistical distribution of \ac{CSI} in a measured dataset.
\acp{GAN} can capture higher-order channel statistics, without requiring manual extraction of parameters, and, by conditioning, can generate plausible \ac{CSI} for arbitrary positions.
However, as presented here, \ac{GAN}-based channel modeling still has many downsides compared to \acp{GSCM}, some important ones being the inferior interpretability, lack of meaningful spatial consistency and limited flexibility of the model:
The channel model is only valid for the particular environment and antenna deployment of the dataset.
For these reasons, we have come to the conclusion that \ac{GAN}-based channel models are not good interpolators and are also not useful as general-purpose \acp{GSCM} in their current form.
Achieving that would likely require more, and more diverse training data and a different architectural approach.
For example we could imagine using \acp{GAN} in conjunction with raytracing-based channel models, where a generative model creates a radio environment, and the raytracer computes spatially consistent \ac{CSI}.

\bibliographystyle{IEEEtran}
\bibliography{IEEEabrv,references}

\end{document}